\documentclass{JHEP3}
\JHEPspecialurl{http://jhep.sissa.it/JOURNAL/JHEP3.tar.gz}
\usepackage{epsfig,multicol,bbm}
\usepackage{amsmath}
\usepackage{placeins}

\newcommand{\be}{\begin{equation}}
\newcommand{\ee}{\end{equation}}
\newcommand{\bea}{\begin{eqnarray}}
\newcommand{\eea}{\end{eqnarray}}

\newcommand{\tev}{\,\, \mathrm{TeV}}
\newcommand{\gev}{\,\, \mathrm{GeV}}

\title{MFV Reductions of MSSM Parameter Space}

\preprint{CERN-PH-TH-2014-234, DAMTP-2014-79}
\author{
S.S.\ AbdusSalam,$^{a,b}$\thanks{Shehu.AbdusSalam@Roma1.infn.it} \hspace{1mm}
C.P.\ Burgess$^{c,d,e}$\thanks{cburgess@perimeterinstitute.ca} \hspace{1mm} and
F.\ Quevedo$^{b,f}$\thanks{F.Quevedo@damtp.cam.ac.uk} \\
$^a$ INFN, Sez. di Roma, P.le A. Moro 2, I-00185 Roma, Italia \\
$^b$ The Abdus Salam ICTP, Trieste, Italy \\
$^c$ Department of Physics \& Astronomy, McMaster University,
Hamilton ON, Canada \\
${}^d$ Perimeter Institute for Theoretical Physics, Waterloo, ON, Canada\\
${}^e$ Division PH\,-TH, CERN, CH-1211, Gen\`eve 23, Suisse\\
$^f$ DAMTP, Cambridge University, Cambridge, UK
}

\abstract{
The 100+ free parameters of the minimal supersymmetric standard model
(MSSM) make it computationally difficult to compare systematically
with data, motivating the study of specific parameter reductions such
as the cMSSM and pMSSM. Here we instead study the reductions of
parameter space implied by using minimal flavour violation (MFV) to
organise the R-parity conserving MSSM, with a view towards
systematically building in constraints on flavour-violating
physics. Within this framework the space of parameters is reduced by
expanding soft supersymmetry-breaking terms in powers of the Cabibbo angle,
leading to a 24-, 30- or 42-parameter framework (which we call
MSSM-24, MSSM-30, and MSSM-42 respectively), depending on the order
kept in the expansion. We provide a Bayesian global fit to data of the
MSSM-30 parameter set to show that this is manageable with current
tools. We compare the MFV reductions to the 19-parameter pMSSM choice
and show that the pMSSM is not contained as a subset. The MSSM-30
analysis favours a relatively lighter TeV-scale pseudoscalar Higgs
boson and $\tan \beta \sim 10$ with multi-TeV sparticles. 
}


\begin{document}

\vspace{1cm}

\begin{quote}
\begin{center}
 {\em Don't panic.} \\
 (on the cover of {\em The Hitchhiker's Guide to the Galaxy} \cite{HHG}) 
\end{center}
\rightline{-- Douglas Adams}
\end{quote}

\vspace{0.1cm}

\section{Introduction}
Supersymmetry, when linearly realised, requires the existence of
superpartners to the known elementary particles, and robustly dictates
their quantum numbers. Less robustly dictated are their masses and
couplings once supersymmetry is spontaneously broken, as experiments
demand it must be. A full description of these requires the more than
100 parameters of the supersymmetry-breaking sector of the R-parity
conserving minimal supersymmetric Standard Model (MSSM).

The challenge of confronting such a vast parameter space with data
drives the development of various kinds of well-motivated benchmark
models. The earliest of these, the cMSSM/mSUGRA \cite{cMSSM},
specialises to a restricted parameter space motivated by what would be
generated if supersymmetry were broken in a flavour-blind hidden
sector (as suggested by the earliest gravity-mediation models). This
simple model is one of the main benchmarks against which LHC results
are compared, with the result that it is in real tension with the
data.

But should this tension be regarded as evidence against supersymmetry,
even if only in its linearly realised\footnote{See \cite{MSLED} for a
  well-motivated example where supersymmetry breaks at the electroweak
  scale but is nonlinearly realised in the Standard Model sector,
  and so {\em doesn't} require the existence of MSSM superpartners
  like squarks and sleptons.}
form? Answering this requires a more detailed exploration of the
parameter space, yet a complete scan of the total parameters still
remains beyond our current computational capabilities. 

What is needed is a more strategic survey of the possibilities, of
which several approaches have emerged. One approach --- for example,
Gauge-Mediated Supersymmetry Breaking (GMSB) \cite{GMSB}, or more
sophisticated string-motivated gravity mediation mechanisms
\cite{StGMM} --- is to explore alternative mechanisms of supersymmetry
breaking whose low-energy implications differ from those of the
minimal gravity-mediated picture. Another focusses less on surveying
the parameter space and more on the generic features of the underlying
production and decay mechanisms, such as appear in `simplified models'
\cite{SimplifiedModels}. Comparison of such models to the data can
quantify which of these mechanisms are favoured or disfavoured. A more
specific `simplified models' approach instead focuses on those
interactions that take part in the naturalness issues that underlie
the motivation for supersymmetry in the first place
\cite{NaturalnessModels}. 

A third approach is to try to broadly survey the allowed parameter
space, but to use prior knowledge about other constraints (like limits
on flavour and CP violations) to cut down the range of parameters
examined at the LHC. Of course this would be simple if it were just a
matter of removing couplings that are excluded by other
constraints. How the parameters are best pruned is more of a judgement
call when the couplings of interest are not directly forbidden by
other observations.

The phenomenological MSSM (pMSSM)\cite{pMSSM} is one of the leading
approaches along these lines which stakes out a 19-parameter subset of 
MSSM by removing all members of potentially dangerous families of
couplings --- such as all flavour-changing interactions beyond those
already in the Standard Model (SM), for example. Besides providing a
good motivation 
for dropping the discarded parameters, the remaining 19-parameter set
is also broad enough to include many models and yet small enough to
allow reasonably systematic comparisons with LHC data. On the other
hand a drawback of the pMSSM is the relatively {\em ad-hoc} way that
the couplings are truncated in detail (in the precise sense described
in more detail below). 

Our goal in this paper is to proceed further along this line of
reasoning, in particular to cast the removal of parameters in terms of
approximate symmetries. This has the advantage of building in at the
outset naturalness constraints since radiative corrections are
guaranteed to respect the choices made for the assumed hierarchies
amongst the model's parameters. In particular we use Minimal Flavour
Violation (MFV) \cite{MFV,MFVSUSY1,MFVBSM,MFVSUSY2} as our main
symmetry criterion to limit flavour-changing physics, wherein the
flavour symmetries of the 
SM in the absence of Yukawa couplings are assumed to be
broken only by other parameters that transform as do the SM Yukawa
couplings themselves. In such a formulation all of the magic of the
GIM mechanism \cite{GIM} is automatically incorporated because
flavour-changing interactions are typically suppressed by the same
small mixing angles as are those of the SM fermions.

When imposed on the MSSM, the MFV hypothesis expresses
flavour-violating supersymmetry-breaking interactions in new basis
which emphasises their transformation properties under the approximate
global flavour symmeties. This makes it possible to associate a power
of the small symmetry-breaking size to all flavour-changing interactions in
a way that is consistent with the known flavour-changes of the SM
itself. Counting the suppression by this symmetry-breaking parameter
provides a natural way to rank their size (and thereby gives a natural
parameter-selection procedure, wherein one neglects all terms beyond a 
fixed order \cite{MFV,MFVSUSY1,Colangelo:2008qp}). Of course, we are
not the 
first to apply MFV methods to models beyond-the-SM (BSM) \cite{MFVBSM}
or to the MSSM \cite{MFVSUSY1,MFVSUSY2}. However, we believe our work
represents the first use of MFV for systematic comparison of the MSSM
to experiment. 

Although very similar in spirit to the ideas behind the pMSSM, the MFV
approach differs in detail and offers several advantages. An important
advantage is the ability to strengthen (or weaken) the MFV
parameter-selection prescription at will to exclude more (or fewer)
interactions, simply by changing the order in symmetry breaking that
is to be neglected. This is to be contrasted with setting all
off-diagonal mass terms to zero by hand once and for all, as is done
for the pMSSM. In this paper we consider three such choices: the
strongest is a 24-parameter MSSM-24, which works at the lowest
nontrivial order. At next order is a 30-parameter MSSM-30 and at the
order beyond this lies a 42-parameter Adams' model,\footnote{Which we
  name in honour of Douglas Adams, who predicted the importance of the
  number 42 for theoretical science \cite{HHG}.}
MSSM-42. Interestingly, none of these parameter sets contain the
pMSSM, which is not defined by any fixed order in the MFV's small
flavour-mixing parameter expansion. 

All three of these are multi-parameter alternatives to the pMSSM. They
contain all of the pMSSM's main virtue and more. They are broad enough
to include a large variety of well-motivated supersymmetric model
points, and yet are small 
enough to bring within reach a systematic comparison with experimental
data. As a first illustration we perform such a comparison for
MSSM-30, showing that even this 30-parameter system is not too large to
be surveyed using reasonable resources. The systematic exploration of
these models should provide a better way for drawing quantitative
inferences regarding whether linearly realised supersymmetry is yet
disfavoured by current data. As is also the case for the pMSSM, the
broader set of parameters contains some atypical expectations
compared to the simpler and more constrained sub-spaces usually
considered, until recently, in supersymmetry searches. Although we
focus only on R-parity invariant interactions, the method can be
easily extended to include the R-parity violating
MSSM~\cite{MFVRPViol}. 

From a bottom-up perspective, the drawback of using {\em ad-hoc}
criteria for reducing the MSSM parameter space is the uncertainty
of the theoretical prejudices that underlie the choices made. 
Selecting to work within a few-parameter framework comes with 
a cost -- a potential loss of physics that may prove important. 
For example, moving from the cMSSM to the 20-parameter pMSSM, as done in
Ref.~\cite{AbdusSalam:2008uv,AbdusSalam:2009qd}, changed the favoured
masses of the Higgs boson and the scalar top-quark to 119-128 GeV
and 2-3 TeV respectively, at a time where such heavy masses were
considered impossible within the traditional cMSSM. Another example:
by setting the off-diagonal mass terms to be zero within the pMSSM
frame, certain diagrams that contribute to flavour changing decays
(such as in the decay $B_s \rightarrow \mu^+ \mu^-$) are lost by
construction. 

The MFV framework provides a natural way to extend the number of 
parameters in a systematic fashion, order-by-order, from the
traditional few-parameters towards the complete and phenomenological
representations. We consider the work we present here as only a first
step towards a more systematic approach to soft terms from a
bottom-up perspective. Jumping from the handful of parameters of the
cMSSM to phenomenological studies of the pMSSM took more than 25 years 
\cite{AbdusSalam:2008uv,Berger:2008cq,AbdusSalam:2009qd} due in part
to the computational challenge of considering more than 5
parameters. Thanks to increasing computing capacities this is becoming
less of an issue. 

The main disadvantage of a Bayesian analysis for models with many
parameters is that prior-dependence can limit the predictive
power. One possible approach in the short term is to seek observables
that are prior-independent and to estimate, qualitatively, the extent
at which current data is able to constrain the supersymmetry
models~\cite{AbdusSalam:2011hd, Sekmen:2011cz,AbdusSalam:2012sy,
  AbdusSalam:2012ir}. In the longer term this is less of an 
issue as better, more constraining, data becomes available. In what
follows we do not explore prior dependence in too much detail, beyond
comparing some of our results with fits to the pMSSM, because our
immediate goal is to define the general set-up for later use. 

Our presentation is organised as follows. We first, in
\S\ref{sec:models} describe in more detail the choices made both in
the pMSSM and in our three realisations of the
MFV-MSSM. \S\ref{sec:fit} then describes a global fit of the MSSM-30
model to the data, with the goal of illustrating the utility of the
MFV approach. Finally \S\ref{sec:concl} briefly summarises our
conclusions.

\section{The models}
\label{sec:models}

In this section we provide a brief summary of the pMSSM and of the
assumptions that go into the MFV-MSSMs that are compared later with
observations. The starting point for both is the observation that a
full comparison of LHC and other experiments data to the 100-plus
parameters of the MSSM is not (yet) feasible, nor is it desirable (at
the moment) given that many of these parameters
describe processes that are strongly constrained by limits on flavour
changing neutral currents (FCNCs) and on CP violation. Therefore we
seek a methodology that allows a maximal probe of the MSSM parameter
space with minimal imposition of {\em ad-hoc} relations or truncations
amongst the free parameters. 

\subsection{Parameter pruning}

We start with a broad-brush description of the pMSSM and MFV-MSSM, in
particular showing how these are related to one another. 

\subsubsection*{The pMSSM}

The goal is to arrive at a criterion for excluding flavour-changing
and CP-violating interactions. The pMSSM does so by making the
following choices \cite{pMSSM}:
\begin{itemize}
\item The absence of flavour-violating interactions (when renormalised
  at TeV scales);
\item Degenerate masses and negligible Yukawa couplings for the first
  two generations of sfermions;
\item No CP-violating interactions (beyond those of the SM CKM
  matrix);
\item $R$-parity conservation;
\item The lightest neutralino should be the lightest superpartner (LSP)
  and a thermal relic.
\end{itemize}

This approach leads to a model for which 19 parameters capture 
superpartner and multiple-Higgs physics. The 19 parameters are: 10
sfermion masses 
($m_{Q_1}$, $m_{Q_3}$, $m_{u_1}$, $m_{u_3}$, $m_{d_1}$, $m_{d_3}$,
$m_{L_1}$, $m_{L_3}$, $m_{e_1}$, $m_{e_3}$); 3 gaugino masses ($M_1$,
$M_2$, $M_3$); 3 trilinear scalar couplings ($A_b$, $A_t$, $A_\tau$);
and 3 Higgs/Higgsino parameters ($\mu$, $M_A$, $\tan\beta$). We see
that its definition includes choices that are well-motivated but
ultimately {\em ad-hoc}. For instance, to avoid extra CP-violating
sources the supersymmetry-breaking terms are set, by hand, to be
real. This amounts to the assertion that no CP-violation 
effects play an important
role in physical processes or interactions at colliders. Similarly,
the first and second generation sfermion masses are set to be
degenerate in order to avoid conflict with the non-observation of
FCNCs, while the flavour changes of the SM are of course kept. It is
necessarily tricky to distinguish BSM physics that explicitly violates
flavour from the higher order corrections through which
flavour-blind BSM physics learns about SM flavour violation.

An alternative approach is to systematically represent {\em
  all}\hspace{1mm} flavour physics effects --- both SM and BSM --- as
a perturbation involving some natural flavour expansion parameter,
such as would be the case in an MFV analysis. Although inspired by MFV
considerations, the pMSSM flavour constraints are {\em not} derived
using MFV symmetry considerations (though this claim is sometimes
made).

\subsubsection*{Minimal flavour violation: the MFV-MSSM}

The MFV hypothesis \cite{MFV} formulates the small size of
flavour-violating effects in terms of approximate symmetries. To this
end the starting point is to 
identify the large group, $G$, of flavour symmetries that the SM enjoys
when all Yukawa couplings vanish. The assumption is then that the only
quantities that break these symmetries are spurion fields that are
proportional to the SM Yukawa couplings themselves. That is, the
action is $G$-invariant when expressed in terms of its regular fields
{\em and} the spurion fields, with the spurion fields then being
replaced by their vacuum expectation values, whose values are inferred
from the SM Yukawa 
couplings. This has the virtue of automatically building in the GIM
cancellations required by observations once loop effects are
included.

As applied to the MSSM the upshot is that MFV boils down to the
requirement that all the low-scale MSSM flavour couplings can be
reconstructed entirely out of  appropriate powers of the SM Yukawa
coupling matrices, $Y_{U,D,E}$, ensuring that flavour violations are
solely governed by the CKM matrix. Within the MFV
framework\footnote{There is also an alternative geometrical approach
  which is not considered here \cite{Ellis:2009di}.}, soft
supersymmetry breaking terms are expanded in series of the
$G$-invariant spurion
factors~\cite{Ellis:2007kb,Colangelo:2008qp,MFV,MFVSUSY1,Smith:2009hj}: 
\bea \label{mfvpars}
& &(M^2_Q)_{ij} = M^2_Q \left[ \delta_{ij} + b_1 (Y_U^\dagger
  Y_U)_{ij} + b_2 (Y_D^\dagger Y_D)_{ij} + c_1\{ (Y_D^\dagger Y_D
  Y_U^\dagger Y_U)_{ij} + H.c.\} + \ldots\right], \nonumber \\ \nonumber
& &(M^2_U)_{ij} = M^2_U \left[ \delta_{ij} + b_3 (Y_U
  Y_U^\dagger)_{ij} + \ldots \right],  \\ \nonumber
& &(M^2_D)_{ij} = M^2_D \left[
  \delta_{ij} + [Y_D (b_6 + b_7 Y_U^\dagger Y_U) Y_D^\dagger]_{ij}
 + \ldots \right],\\
& &(M^2_L)_{ij} = M^2_L \left[ \delta_{ij} + b_{13} (Y_E^\dagger
  Y_E)_{ij} + \ldots \right], \\ \nonumber
& &(M^2_E)_{ij} = M^2_E \left[ \delta_{ij} +
  b_{14} (Y_E Y_E^\dagger)_{ij} + \ldots \right] \,,
\eea
and
\bea \label{mfvpars2}
& &(A^{'}_E)_{ij} = a_E \left[ \delta_{ij} + b_{15} (Y_E^\dagger
  Y_E)_{ij} + \ldots \right], \nonumber\\ 
& &(A^{'}_U)_{ij} = a_U \left[ \delta_{ij} + b_9 (Y_U^\dagger
  Y_U)_{ij} + b_{10} (Y_D^\dagger Y_D)_{ij} + \ldots \right], \\
& &(A^{'}_D)_{ij} = a_D \left[ \delta_{ij} + b_{11} (Y_U^\dagger
  Y_U)_{ij} + b_{12} (Y_D^\dagger Y_D)_{ij} + c_6 (Y_D^\dagger Y_D
  Y_U^\dagger Y_U)_{ij} + \ldots \right].\nonumber
\eea
Although the ellipses appear to denote an infinite series, this
collapse to only a few terms due to the Cayley-Hamilton identities for
$3 \times 3$ matrices. For instance, any generic matrix can be written
in the form in Eq.(\ref{mfvpars}), but generically the required
coefficients, $b_i$ and $c_i$, would span many orders of 
magnitude. The power of the MFV hypothesis lies in the assumption that
the $b_i$ and $c_i$ are of order unity, with all small numbers
suppressing flavour changes coming solely from those already in the
Yukawa matrices. Trilinear scalar couplings similarly take the form
$(A_{E,U,D})_{ij} = (A^{'}_{E,U,D} Y_{E,U,D})_{ij}$. 

Now, a non-symmetry way to truncate the above parameters to a 
flavour-blind set is to impose $b_i = c_i =0$. This sets all
off-diagonal elements of the matrices to zero and all diagonal
elements are set to be equal to one another, leading to  
a 14-parameter flavour-blind MSSM with no extra-SM sources of CP
violation. Note these choices ensure the sfermion masses within each
family are degenerate. Lifting the degeneracy to only the first two
generations then gives the 19-parameter pMSSM. This shows how the
pMSSM is related to the MFV MSSM, and why some of the assumptions in
its construction do not rely on symmetries. By contrast, the number of
MFV MSSM parameters in principle is the same as for the original MSSM,
if we work to all orders in the small Yukawa couplings. However,
within the MFV MSSM the number of parameters can be reduced in a
systematic way by dropping terms smaller than a particular fixed order
in small mixing angles (like the Cabibbo angle), as we now see.

\subsection{Expansions in small mixing angles}

A systematic approach for selecting the number of MSSM parameters have 
been prescribed in Ref.~\cite{Colangelo:2008qp}. The counting rule 
explores the hierarchical structure along the off-diagonals terms of
the Yukawa 
matrices usually expressed in terms of the Cabibbo angle, $\lambda = 
\sin\theta_{CB} \simeq 0.23$. The idea starts from the observation
that 
after the collapse of the infinite series in Eq.(\ref{mfvpars}) and
Eq.(\ref{mfvpars2}) into  
few terms by employing the Cayley-Hamilton identities, large pieces of
the terms such as $(Y_U^\dagger Y_U)_{ij}^2$ and $(Y_U^\dagger Y_U)_{ij}$
are proportional to $V_{3i}^*V_{3j}$ where $V$ is the CKM matrix. The
next relatively smaller terms are proportional to $V_{2i}^*
V_{2j}$. So $V_{3i}^* V_{3j}$ and $V_{2i}^* V_{2j}$ can be used as
basis vectors with coefficients of order one and $y_c^2 \sim
\lambda^8$ respectively instead of $(Y_U^\dagger Y_U)_{ij}^2$ and
$(Y_U^\dagger Y_U)_{ij}$. Similarly, instead of $(Y_D^\dagger
Y_D)_{ij}$ and $(Y_D^\dagger Y_D)_{ij}^2$,  $\delta_{i3}^*
\delta_{j3}$ and $\delta_{i2}^* \delta_{j2}$ can  be used with order
$y_b^2$ and $y_s^2$ coefficients respectively. Here $\delta_{ij}$ is
the unit matrix in family space. This way, all possible
multipliable structures lead to new complete basis vectors that
form a closed algebra under multiplication:
\begin{eqnarray}
  \label{xbasis}
  \begin{array}{cccc}
    X_1 = \delta_{3i} \delta_{3j}, & X_2 = \delta_{2i} \delta_{2j}, &
    X_3 = \delta_{3i} \delta_{2j}, & X_4 = \delta_{2i} \delta_{3j},
    \\
    X_5 = \delta_{3i} V_{3j}, & X_6 = \delta_{2i} V_{2j}, & X_7 =
    \delta_{3i} V_{2j}, &X_8 = \delta_{2i} V_{3j}, \\
    X_9 = V^*_{3i} \delta_{3j}, & X_{10} = V^*_{2i} \delta_{2j}, &
    X_{11} = V^*_{3i} \delta_{2j}, & X_{12} = V^*_{2i} \delta_{3j},
    \\
    X_{13} = V^*_{3i} V_{3j}, & X_{14} = V^*_{2i} V_{2j}, & X_{15} =
    V^*_{3i} V_{2j}, & X_{16} = V^*_{2i} V_{3j}.
  \end{array}
\end{eqnarray}
Note that the basis vectors are all of order one since each has at least
one entry of order unity. With these, each of the MFV
parameters can be assigned an order in $\lambda$. Once the accuracy of
calculations is chosen in the form ${\cal O}(\lambda^n)$, then the
prescription can be used to systematically discard terms within the
supersymmetry-breaking parameters expansion expressed in the $X_i$ basis.

\subsubsection*{The MSSM-42 model}

For instance, as done in Ref.~\cite{Colangelo:2008qp}, dropping terms
of order $\lambda^6 \sim 10^{-4}$ and higher from the soft
supersymmetry-breaking terms in Eq.(\ref{mfvpars}) and
Eq.(\ref{mfvpars2}), the MSSM parameters become: 
\bea \label{mfvparsx}
e^{\phi_1} M_1, & &\,\quad  e^{\phi_2} M_2, \,\quad M_3, \,\quad
 e^{\phi_\mu} \mu, \,\quad M_A, \,\quad \tan \beta\nonumber \\ \nonumber
 M^2_Q &=& \tilde{a}_1 + x_1 X_{13} + y_1 X_1 + y_2 X_5 + y^*_2 X_9,\\\nonumber
 M^2_U &=& \tilde{a}_2  + x_2 X_1, \\\nonumber
 M^2_D &=& \tilde{a}_3 + y_3 X_1 + w_1 X_3 + w^*_1 X_4, \\
 M^2_L &=& \tilde{a}_6 + y_6 X_1, \\\nonumber
 M^2_E &=& \tilde{a}_7 + y_7 X_1, \\\nonumber
 A_E &=& \tilde{a}_8 X_1 + w_5 X_2, \\\nonumber
 A_U &=& \tilde{a}_4 X_5 + y_4 X_1 + w_2 X_6, \\\nonumber
 A_D &=& \tilde{a}_5 X_1 + y_5 X_5 + w_3 X_2 + w_4 X_4.\nonumber
\eea
Since the squark supersymmetry-breaking mass parameters are Hermitian then
$\tilde{a}_{1-3,6,7} > 0, x_1, x_2, y_1, y_3, y_6, y_7$ must be real
while the other coefficients can be complex. Hence the total number of
supersymmetry-breaking parameters amounts to 42, defining the MSSM-42.

\subsubsection*{The MSSM-30 model}
Alternatively, keeping only those terms of order ${\cal O}(\lambda^4)
\sim {\cal O}(10^{-3})$ means keeping only $x_{1-2}, y_1, y_3,
y_6, y_7 \in \mathbb{R}$; and $\tilde{a}_{4,5,8}, y_{4-5} \in
\mathbb{C}$ terms from Eq.(\ref{mfvparsx}). These make a 30-parameters
MSSM-30:
\bea \label{mfvpar30}
& &e^{\phi_1} M_1, \,\quad e^{\phi_2} M_2, \,\quad M_3, \,\quad
\mu, \,\quad M_A, \,\quad \tan \beta, \,\quad e^{\phi_\mu}, \nonumber\\ \nonumber
& &M^2_Q = \tilde{a}_1 + x_1 X_{13} + y_1 X_1,\\ \nonumber
& &M^2_U = \tilde{a}_2  + x_2 X_1, \\\nonumber
& &M^2_D = \tilde{a}_3 + y_3 X_1, \\
& &M^2_L = \tilde{a}_6 + y_6 X_1, \\\nonumber
& &M^2_E = \tilde{a}_7 + y_7 X_1, \\\nonumber
& &A_E = \tilde{a}_8 X_1, \\\nonumber
& &A_U = \tilde{a}_4 X_5 + y_4 X_1, \\\nonumber
& &A_D = \tilde{a}_5 X_1 + y_5 X_5
\eea

\subsubsection*{The MSSM-24 model}
Going doing to ${\cal O}(\lambda^3) \sim {\cal O}(10^{-2})$, only
$x_{1-2} \in \mathbb{R}$; and $y_5 \in \mathbb{C}$ remain from the
non-diagonal mass and trilinear coupling expansion terms in
Eq.(\ref{mfvpar30}). These make a total of 24 soft-supersymmetry breaking
parameters for MSSM-24:
\bea \label{mfvpar24}
& &e^{\phi_1} M_1, \,\quad e^{\phi_2} M_2, \,\quad M_3, \,\quad
\mu, \,\quad M_A, \,\quad \tan \beta, \,\quad e^{\phi_\mu},\nonumber \\ \nonumber
& &M^2_Q = \tilde{a}_1 + x_1 X_{13},\\ \nonumber
& &M^2_U = \tilde{a}_2  + x_2 X_1, \\\nonumber
& &M^2_D = \tilde{a}_3, \\
& &M^2_L = \tilde{a}_6, \\\nonumber
& &M^2_E = \tilde{a}_7, \\\nonumber
& &A_E = \tilde{a}_8 X_1, \\\nonumber
& &A_U = \tilde{a}_4 X_5, \\\nonumber
& &A_D = \tilde{a}_5 X_1 + y_5 X_5.
\eea
Note that the MFV MSSM parametrisation cannot be reduced to the 19
parameters of the pMSSM.

\subsubsection*{The MSSM-11 model}

Ideally we would like to reduce the number of parameters even further, 
keeping the systematic approach we are following here. This we cannot
do, but it is possible to define a minimal extension of the
constrained MSSM --- {\em i.e.} the cMSSM --- in a more {\em ad hoc}
way by setting in the above:
$\phi_1=\phi_2=0$, $\tilde{a}_1 = \tilde{a}_2 = \tilde{a}_3 =
\tilde{a}_6 = \tilde{a}_7 =
m_0$, $Re(\tilde{a}_8) = Re(\tilde{a}_4) = Re(\tilde{a}_5)= A_0$, and
$Im(\tilde{a}_8) = Im(\tilde{a}_4) = Im(\tilde{a}_5)= 0$. That is,
\bea \label{cmssm9}
 & &M_1= M_2=M_3=m_{1/2}, \,\quad Im(m_{1/2}), \\ \nonumber
 & &\{\mu, M_A,
  e^{\phi_\mu}\} \rightarrow \{m_{H_1}=m_{H_2}=m_0, sign(\mu)\}, \,\quad
 \tan \beta, \\ \nonumber
 & &M^2_Q = m_0^2 + x_1 X_{13}, \,\quad M^2_U = m_0^2  + x_2 X_1,
 \,\quad  M^2_D = M^2_L = M^2_E = m_0^2, \,\quad Im(m_0) \\\nonumber
 & &A_E = A_0 X_1, \,\quad A_U = A_0 X_5, \,\quad  A_D = A_0 X_1
 + y_5 X_5, \, \quad Im(A_0, \, y_5).
\eea
This reduces the parameter space into an 11-parameters cMSSM or
cMSSM-11. Given its simplicity it may be worth studying this model in
detail even though it reintroduces some {\it ad-hoc} selection of
parameters at the end.

Out of these sub-MSSMs derived via the MFV MSSM scheme, in this paper
we concentrate on the MSSM-30 model and fit its parameters to
experiments data as a first step into landscaping, and making further
forecasts about, the MSSM parameter space.

\section{The MSSM-30 fit}
\label{sec:fit}
As mentioned earlier the MSSM-42 model cannot be reduced into the
traditional pMSSM parameter space. The MSSM-24 is the closest
to the pMSSM. However, looking at the parameter lists of the
MSSM models mentioned in the previous section we select MSSM-30 for
going beyond the pMSSM especially in the flavour sector. This is a
first-step beyond the pMSSM within our series of
MSSM projects \cite{Feroz:2008wr, AbdusSalam:2008uv,
  AbdusSalam:2009qd, AbdusSalam:2009tr,
  AbdusSalam:2010qp, AbdusSalam:2011hd, AbdusSalam:2011fc,
  AbdusSalam:2012sy, AbdusSalam:2012ir, AbdusSalam:2013qba} that
is systematically built for absorbing experimental data
from both energy- and intensity frontiers to high-energy physics
explorations. The explorations of the MSSM CP-violating phases within
various constructs can be found in the literature such as in
Refs.~\cite{Bartl:2001wc,Ellis:2007kb,Colangelo:2008qp,Mercolli:2009ns,Berger:2013zca}.
The sub-MSSMs mainly fall into one of the the various constrained MSSMs,
the pMSSM or flavour-blind MSSM with variable extra-SM CP phases. The MSSM-30
goes beyond these by construction, considering that the systematic
inclusion of the flavour-violating terms is important, and by number
of parameters. The procedure for the Bayesian fit of the MSSM-30 to
data is described as follows.

\subsection{Fitting procedure}
We use Bayesian statistical methods for fitting the MSSM-30 to
data. Bayes' 
theorem takes two input information for deriving essentially two
inference about the model addressed. The process has to be within a
well-defined context. The context, $\cal{H}$, for the MSSM-30 analysis
is that the model represents R-parity preserving linearly realised
supersymmetry and that the neutralino LSP make at least part of the
cold dark matter (CDM) relic. One of the input is the assumption about
the nature of the
model parameters, $\underline{\theta}$. Here we assume a flat prior
probability density, $p(\underline \theta | {\cal{H}})$, over the
MSSM-30 parameters in Eq.(\ref{mfvpar30})
\bea \label{30parameters}
\underline{\theta} \equiv \{ \, & &
M_{1,2,3}, \,\quad \mu, \,\quad M_A, \,\quad \tan \beta, \,\quad
Im(M_{1,2}, \,\quad \mu), \,\quad \tilde{a}_{1,2,\ldots,8}, \\\nonumber
& & Im(\tilde{a}_{4,5,8}), \,\quad x_{1,2}, \,\quad y_{1,3,4,5,6,7},
\,\quad Im(y_{4,5}) \,\quad \}
\eea
where $M_1$, $M_2$ and $M_3$ are the gaugino mass parameters varied
in the range -4 to 4 TeV for both real and imaginary parts of
$M_{1,\, 2}$ and 100 GeV to 4 TeV for $M_3$. The mass-term parameters
$\tilde{a}_{1,2,3,6,7} > 0$ are
varied within the range $(100 \gev)^2$ to $(4 \tev)^2$ and $-(4
\tev)^2$ to $(4 \tev)^2$ for $x_{1,2}, y_{1,3,6,7}$. The trilinear
coupling terms $\tilde{a}_{4,5,8}, Im(\tilde{a}_{4,5,8}), y_{4,5}$,
and $Im(y_{4,5})$ are varied within $-8 \tev$ to $8 \tev$. The
Higgs-sector parameters are specified by the speudoscalar Higgs masses $M_A$,
varied between $100 \gev$ to $4 \tev$ and the Higgs doublets mixing term
$\mu, \, Im(\mu)$ both varied within the range -4 to 4 TeV.
The ratio of the vacuum expectation values $\tan
\beta=\left<H_2\right>/\left<H_1\right>$  is allowed to be between 2
and 60 \footnote{The $4 \tev$ range is taken having the $14 \tev$ LHC
  capabilities in mind and also for consistency reasons (as well for
  the $\tan \beta$ range) to allow possible comparisons with our
  previous pMSSM work.}. The SM parameters are fixed according to
experimental results as: mass of the Z-boson, $m_Z = 91.2 \gev$, 
top quark mass, $m_t = 165.4 \gev$, bottom quark mass, $m_b = 4.2
\gev$, the electromagnetic coupling, $\alpha_{em}^{-1} = 127.9$, and
the strong interaction coupling, $\alpha_s = 0.119$.
\TABLE[t]{\small%
\begin{tabular}{|ll||ll|}\hline
Observable & Constraint & Observable & Constraint  \\
\hline
$m_W$ [GeV]& $80.399 \pm  0.023$ \cite{verzo}&$A^l = A^e$& $0.1513 \pm
0.0021$ \cite{:2005ema} \\
$\Gamma_Z$ [GeV]& $2.4952 \pm 0.0023$ \cite{:2005ema}&$A^b$ & $0.923
\pm 0.020$ \cite{:2005ema}\\
$\sin^2\, \theta_{eff}^{lep}$  & $0.2324 \pm 0.0012$ \cite{:2005ema}&$A^c$ & $0.670 \pm 0.027$ \cite{:2005ema}\\
$R_l^0$ & $20.767 \pm 0.025$ \cite{:2005ema} &$Br(B_s \rightarrow
\mu^+ \mu^-)$ & $3.2^{+1.5}_{-1.2} \times 10^{-9}$
\cite{Aaij:2012nna}\\
$R_b^0$ & $0.21629 \pm 0.00066$ \cite{:2005ema}&$\Delta M_{B_s}$ &
$17.77 \pm 0.12$  ps$^{-1}$ \cite{Abulencia:2006ze}\\
$R_c^0$ & $0.1721 \pm 0.0030$ \cite{:2005ema}&$R_{Br(B_u \rightarrow \tau \nu)}$&
$1.49 \pm 0.3091$ \cite{Aubert:2004kz}\\%paoti,hep-lat/0507015}\\
$A_{\textrm{FB}}^b$ & $0.0992 \pm 0.0016$ \cite{:2005ema}& $\Delta
M_{B_d}$ & $0.507 \pm 0.005$ ps$^{-1}$\cite{Barberio:2008fa} \\
$A_{\textrm{FB}}^c$ & $0.0707 \pm 0.0035$ \cite{:2005ema}&$\Omega_{CDM} h^2$ & $0.11
\pm 0.02 $ \cite{0803.0547}\\
$m_h$ [GeV] & $125.6 \pm 3.0$ \cite{ATLAS:2013mma, CMS:yva} &  $Br(B_d
\rightarrow \mu^+ \mu^-)$ & $<1.8 \times 10^{-8}$ \cite{Aaij:2011rja}\\
$d_\mu$ &$<2.8\times 10^{-19}$ \cite{McNabb:2004tj}& $Br(B\rightarrow
X_s \gamma)$ & $(3.52 \pm 0.25) \times 10^{-4}$ \cite{Barberio:2007cr}\\
$d_\tau$ &$<1.1\times 10^{-17}$ \cite{Nakamura:2010zzi}
&$d_e$ &$<1.6\times 10^{-27}$ \cite{Regan:2002ta}\\
\hline
\end{tabular}
\caption{Summary for the central values and errors for the electroweak
  physics, B-physics, dipole moments of the leptons and cold dark
  matter relic density constraints. \label{tab:obs}}
}

The second input within Bayes' theorem is the data, $\underline
d$. The data used for fitting 
the MSSM-30 are summarised in Table~\ref{tab:obs}. It is made of the
experimental central values ($\mu_i$) 
and errors ($\sigma_i$) for the Higgs boson mass, the electroweak
physics, B-physics, dipole moments of the leptons and the CDM relic
density observables listed in the set $\underline O$:
\bea
\underline O &\equiv &\{ m_h, \; m_W,\; \Gamma_Z,\; \sin^2\,
\theta^{lep}_{eff},\; R_l^0,\; R_{b,c}^0,\;
A_{FB}^{b,c},\;  A^l = A^e,\; A^{b,c}, \\ \nonumber
& &BR(B \rightarrow X_s \, \gamma),\; BR(B_s \rightarrow \mu^+ \,
\mu^-),\; \Delta M_{B_s},\;  R_{BR(B_u \rightarrow \tau \nu)},\\ \nonumber
& &\Omega_{CDM}h^2, \; Br(B_d \rightarrow \mu^+ \mu^-), \; \Delta
M_{B_d}, \; d_{e,\mu,\tau} \}.
\eea
For this analysis, the constraint from anomalous magnetic
dipole moment of the muon is not included in order to avoid possible
tension with the EDM constraints since this have
the potential of slowing down the exploration of the MSSM-30 parameter
space. 

The compatibility of the MSSM-30 with the data is quantified at each
point in parameter space by the likelihood, 
the probability of the data set given the parameter point,
$p(\underline d|\underline \theta, {\cal{H}})$. Assuming the
observables are independent\footnote{We did not add the experimental
  correlations between some of the electroweak observables. This
  simplification is plausible since they are rather mildly
  constraining.}, the combined likelihood
\be p(\underline d|\underline \theta, {\cal{H}}) = L(x) \,
\prod_i \, \frac{ \exp\left[- (O_i - \mu_i)^2/2
    \sigma_i^2\right]}{\sqrt{2\pi \sigma_i^2}}
\ee where the index $i$ runs over the list of observables $\underline
O$, the variable $x$ represents the predicted value of neutralino CDM
relic density at an MSSM-30 parameter point and
\be \label{olik}
L(x) =
\begin{cases}
  1/(y + \sqrt{\pi s^2/2}) &  \textrm{ if } x < y \\
  \exp\left[-(x-y)^2/2s^2\right]/(y + \sqrt{\pi s^2/2}) &
  \textrm{ if } x \geq y
\end{cases}.
\ee
Here $y = 0.11$ is the CDM relic density central value and $s=0.02$
the corresponding inflated (to allow for theoretical uncertainties)
error.

The MSSM-30 parameters are passed to {\sc
  SPHENO}~\cite{Porod:2003um, Porod:2011nf} and {\sc
  SUSY\_FLAVOR}~\cite{Crivellin:2012jv} packages~\footnote{Note that
  in this work no vacua analyses \cite{Camargo-Molina:2013qva, Camargo-Molina:2014pwa} were carried out beyond those implemented in the spectrum calculators.}, via the
SLHA2~\cite{Allanach:2008qq} interface, for computing the supersymmetry
spectrum, mixing angles and couplings; and corresponding predictions:
the branching ratios $BR(B_s \rightarrow \mu^+ \mu^-)$, $BR(B
\rightarrow s \gamma)$, $R_{BR(B_u
  \rightarrow \tau \nu)}$, $Br(B_d \rightarrow \mu^+ \mu^-)$, $\Delta
M_{B_s}$,  $\Delta M_{B_d}$ and $d_{e,\mu,\tau}$.
Using the SLHA1~\cite{Skands:2003cj} interface, the neutralino CDM relic density
were computed using {\sc micrOMEGAs}~\cite{Belanger:2008sj} while
{\sc susyPOPE}~\cite{Heinemeyer:2006px,Heinemeyer:2007bw}
is used for computing precision observables that include the $W$-boson
mass $m_W$, the effective leptonic mixing angle variable $\sin^2
\theta^{lep}_{eff}$, the total $Z$-boson decay width, $\Gamma_Z$, and
the other electroweak observables whose experimentally
determined central values and associated errors are summarised in
Table~\ref{tab:obs}. The predictions from {\sc SUSY\_FLAVOR} were not
used for fitting the MSSM-30 but could be used for comparing
predictions from the two packages. The {\sc
  MultiNest}~\cite{Feroz:2007kg,Feroz:2008xx} package which implements
the Nested Sampling algorithm~\cite{Skilling} were used for fitting
the MSSM-30
to data. The results of the Bayesian fit are the posterior probability
density of the model parameters given the data, $p(\underline \theta|
\underline d,{\cal{H}})$, and the support (or evidence), ${\cal Z} =
p(\underline d|{\cal{H}})$, for the MSSM-30 from the data 
used. These come directly from Bayes' theorem
\be \label{bayes}
p(\underline \theta |\underline d, {\cal{H}}) \, \times \,
p(\underline d|{\cal{H}}) = p(\underline d|\underline
\theta,{\cal{H}}) \, \times \, p(\underline \theta|{\cal{H}}).
\ee
The posterior probability densities of the MSSM-30 parameters and
representative sparticle masses are presented in the next
subsection.

\subsection{Posterior distributions}
The quantities of interest to be investigated from the output of the
MSSM-30 fit to data are the supersymmetry-breaking parameters and the sparticle
masses. The former provide an indication of the preferred regions
within the MSSM-30 hyperspace which are compatible with the
experimental results while from the latter an insight could be
obtained concerning the prospects for detecting sparticles at the LHC
and/or future colliders.

The one-dimensional posterior probability distributions for the
MSSM-30 parameters are shown in Figure~\ref{fig:30pars}. The real
and imaginary parts of the complex parameters are plotted on the same
figure while the corresponding magnitudes and phases are shown in
Figure~\ref{fig:par30II}. In addition we also present, in
Figure~\ref{fig:par30II}, the posterior distribution for the nature
of the neutralino LSP's gaugino-Higgsino composition $(1 - Z_g)$ where
$Z_g = |N_{11}|^2 + |N_{22}|^2$ with an LSP bino $\tilde{b}$, wino
$\tilde{w}^3$ and Higgsinos $\tilde{H}_{1,2}$ combination
\be \label{nis}
\tilde{\chi}_1^0 = N_{11}\tilde{b} + N_{12}\tilde{w}^3 +
   N_{13}\tilde{{{H}}_1^0} + N_{14}\tilde{H_2^0}, \quad \sum_{i=1,2,3,4}
   (N_{1i})^2 = 1.
\ee
$N_{1i}$ with $i=1,2,3,4$ are coefficient depending on the
supersymmetry-breaking parameters~\cite{ElKheishen:1992yv}. The neutralino is
dominantly Higgsino- or gaugino-like for $(1 - Z_g)$ approximately
equal to unity or zero respectively. The nature of the LSP composition
in relevant for understanding the posterior distributions of the
gauge-sector supersymmetry-breaking parameters.

From Figure~\ref{fig:par30II}, it can be seen that the LSP and
lightest chargino are quasi-degenerate, $m_{\chi_1^\pm} \sim
m_{\chi_1^0} \sim \mu$. Secondly, the posterior of distribution of $(1
- Z_g)$ indicates that the LSP is mostly higgsino-like. Therefore
there is an
efficient neutralino-chargino co-annihilations taking place for
satisfying the CDM relic density requirement. The posterior
distributions for the gaugino mass parameters $M_1$ and $M_2$, and the 
electroweak symmetry breaking constraint control the nature of the
neutralino gaugino­higgsino admixtures. $M_1$ and $M_2$ remain 
approximately unconstrained because the scenario is similar to the
cMSSM's focus point region, see e.g. \cite{Feng:1999zg}, where the
renormalisation group running of $m_{H_2}$ is decoupled from the
gaugino and trilinear parameters. This is the case for the pMSSM
distributions shown in dashed lines except for $M_2$ which looks quite
different apparently due to the non-negligible interplay of the EDMs
and other constrains on the imaginary parts $Im(M_1)$ and $Im(M_2)$ as
shown in Figure~\ref{fig:30pars} or the corresponding phases
($\phi_{1,2}$) shown in Figure~\ref{fig:par30II}.
 The gluino mass distribution is slightly preferred to be heavier
 relative to that in \cite{AbdusSalam:2008uv,AbdusSalam:2009qd} due
 to the intensity-frontier constraints.

Unlike the case for the gluino mass,
the posterior distributions in Figure~\ref{fig:30pars} show that
the intensity-frontier constraints, plus fixing $m_h = 125 \pm 3 \gev$
favour smaller values of $\tan 
\beta$ and lighter pseudoscalar Higgs boson mass $M_A$ relative to the
fits in \cite{AbdusSalam:2008uv,AbdusSalam:2009qd}. The $\tan
\beta$ feature together with the
tendencies for heavier gluinos and sparticles are compatible
with the effect of the EDM constraints. The EDMs tend to be
proportional to $\tan \beta$~\cite{Pospelov:2005pr} so the prevention
of EDM over-production necessarily requires lower $\tan \beta$
values. The fit
indicates a $95\%$ credible interval (Bayesian confidence interval) of
$4.5$ to $26.9$ with a mean value of $\langle \tan \beta \rangle = 13.4 \pm
5.8$. The value of $M_A$ is within the range of 327.8 to 3803.3 GeV
at $95\%$ credible interval with a mean-value $\langle M_A \rangle = 1751.9 \pm 1024.5
\gev$. The application of the ATLAS and CMS collaborations' search for
MSSM Higgs bosons results~\cite{Aad:2014vgg,Khachatryan:2014wca}
on the MSSM-30 posterior would require a dedicated 
interpretation of their data within the new MSSM frame. 

The remaining MSSM-30 parameters, which appear in the mass-squared 
terms, $\tilde{a}_{1,2,3,6,7}$, $x_{1,2}$, $y_{1,3,6,7}$ and those
that appear in the trilinear couplings, $\tilde{a}_{4,5,8}, y_{4,5}$ 
cannot be compared due to their absence within the pMSSM. However, the
mass-squared terms can be compared as shown in 
Figure~\ref{fig:par30II}. It can be seen that the posterior sample
from the flat-prior fit of the MSSM-30 to data favours
supersymmetry-breaking parameters in regions deeper into the multi-TeV
scale beyond the pMSSM results. This feature is expected for scenarios
that alleviates the supersymmetry CP problems (see
Refs.~\cite{Cohen:1996vb,Pospelov:2005pr,Paradisi:2009ey} and
references therein, for instance). The CP-violating phases have to be
either small or the sparticles be heavy into the multi-TeV
regions. The phases were not restricted to be small for the MSSM-30
fit. The feature is also supported by the fact that radiative
corrections to the lightest CP-even Higgs boson mass require heavy 3rd
generation squarks for making up the constraint $m_h = 125 \pm 3
\gev$. The trilinear
couplings, on another hand, are peaked around zero because values
away tend to solutions with negative squark masses. The only exception
here is for the leading parameter ($a_4$, $Im(a_4)$ in
Figure~\ref{fig:30pars} and the corresponding magnitude $a_4$ and phase
$\phi_{a_4}$ shown in Figure~\ref{fig:par30II}) in the trilinear
coupling term $A_U$ which is roughly fixed by the $m_h = 125 \gev$
constrain.

\FIGURE{
\epsfig{file=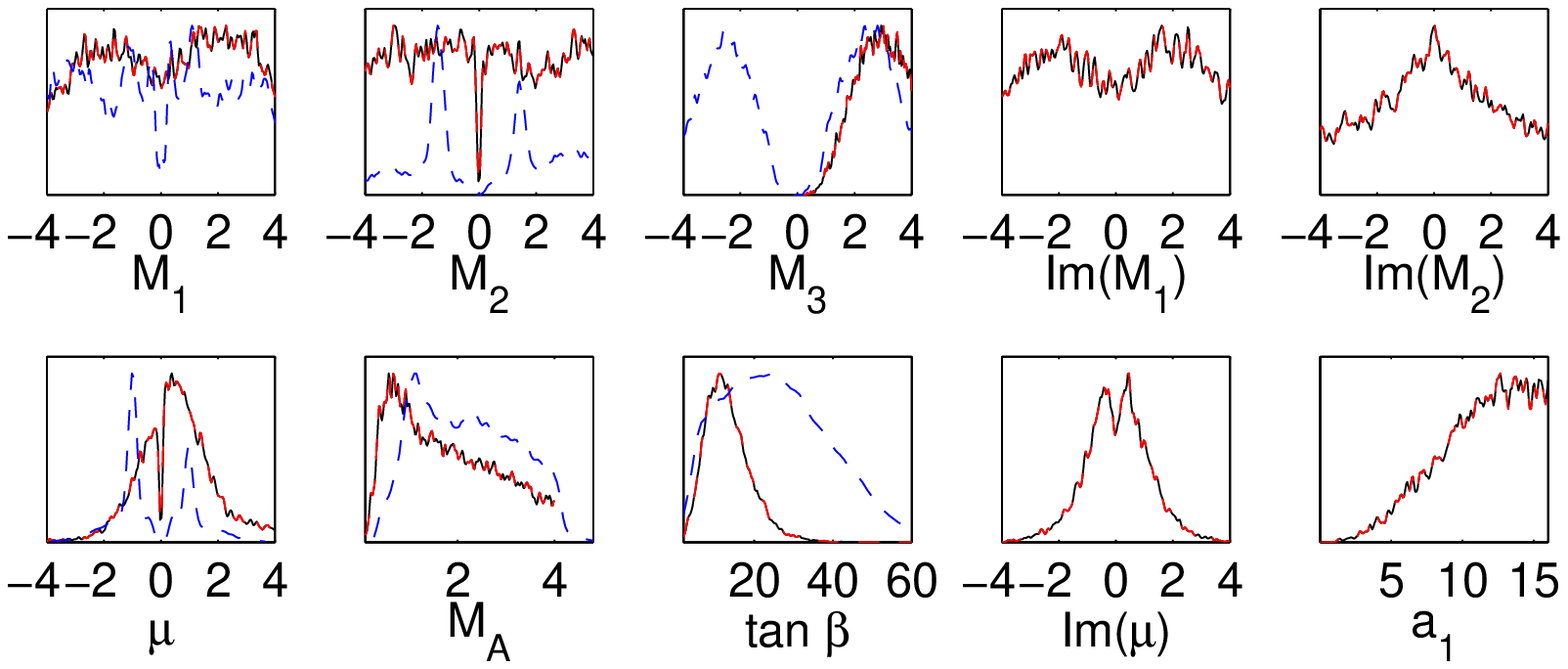,width=.92\textwidth}
\epsfig{file=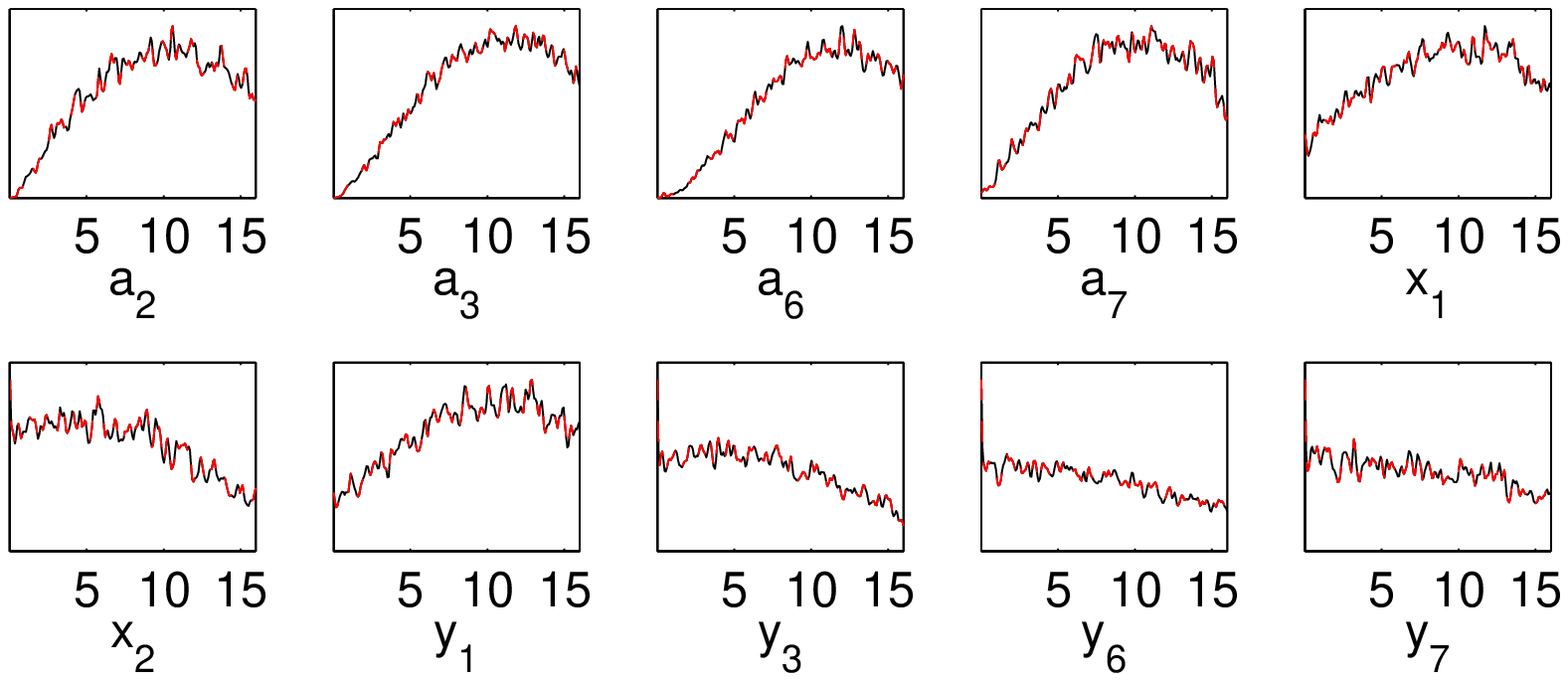,width=.92\textwidth}
\epsfig{file=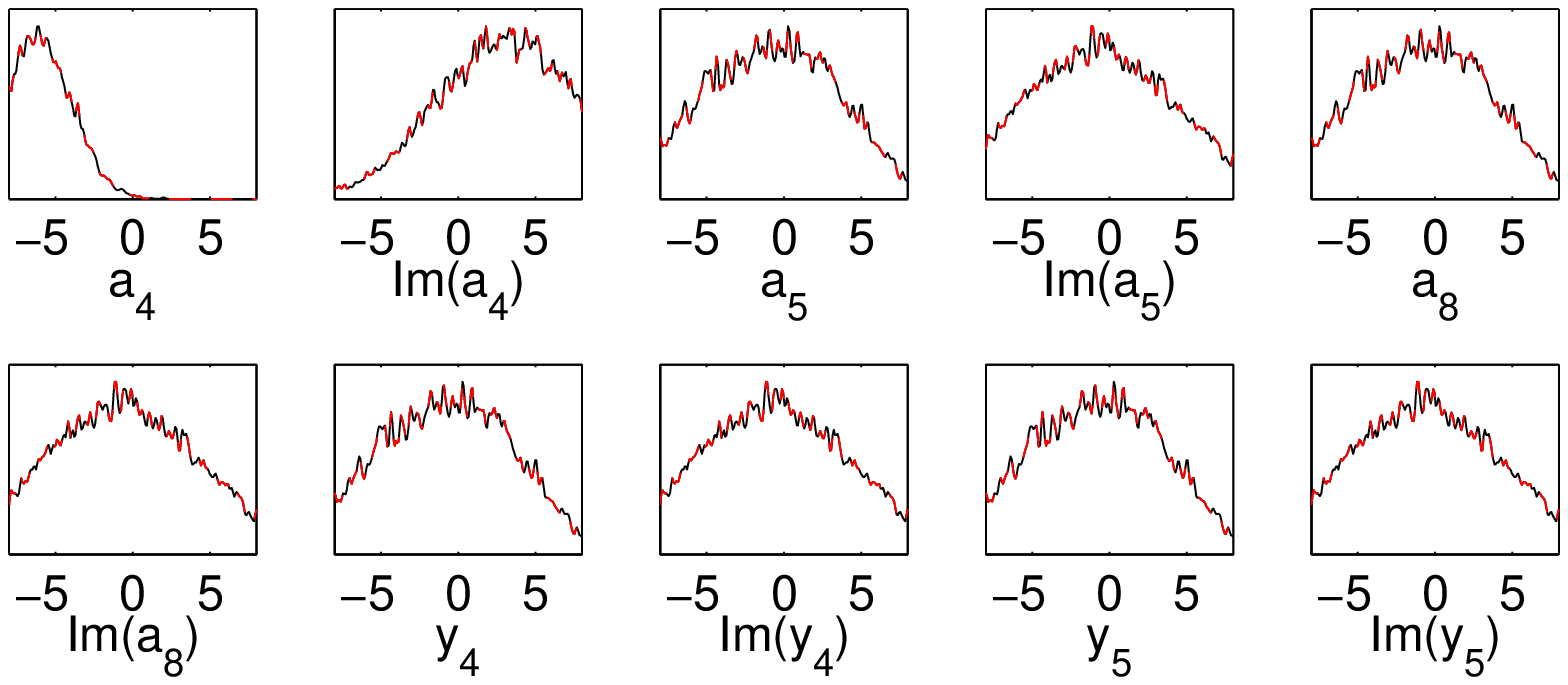,width=.92\textwidth}
\caption{The plots show the posterior probability density functions,
  shown with solid curves, for the MSSM-30 parameters based on flat
  prior fit to the data in Table~\ref{tab:obs}. The dashed curves
  represent the corresponding posteriors for the 2008/9 pMSSM
  fits \cite{AbdusSalam:2008uv,AbdusSalam:2009qd} (when these are available) for
  comparison with the current MSSM-30 fit.  The axes
  $M_{1,2,3}, \, \mu, \, M_A, \,
  Im(M_{1,2}, \, \mu)$ are in $\tev$ units. The parameters in the
  mass-squared terms, $\tilde{a}_{1,2,3,6,7}, \, x_{1,2}, y_{1,3,6,7}$
  are in $\tev^2$. The parameters entering the trilinear couplings
  $\tilde{a}_{4,5,8}, y_{4,5}$ are in $\tev$ units.}
\label{fig:30pars}
}
\FloatBarrier

\FIGURE{
\epsfig{file=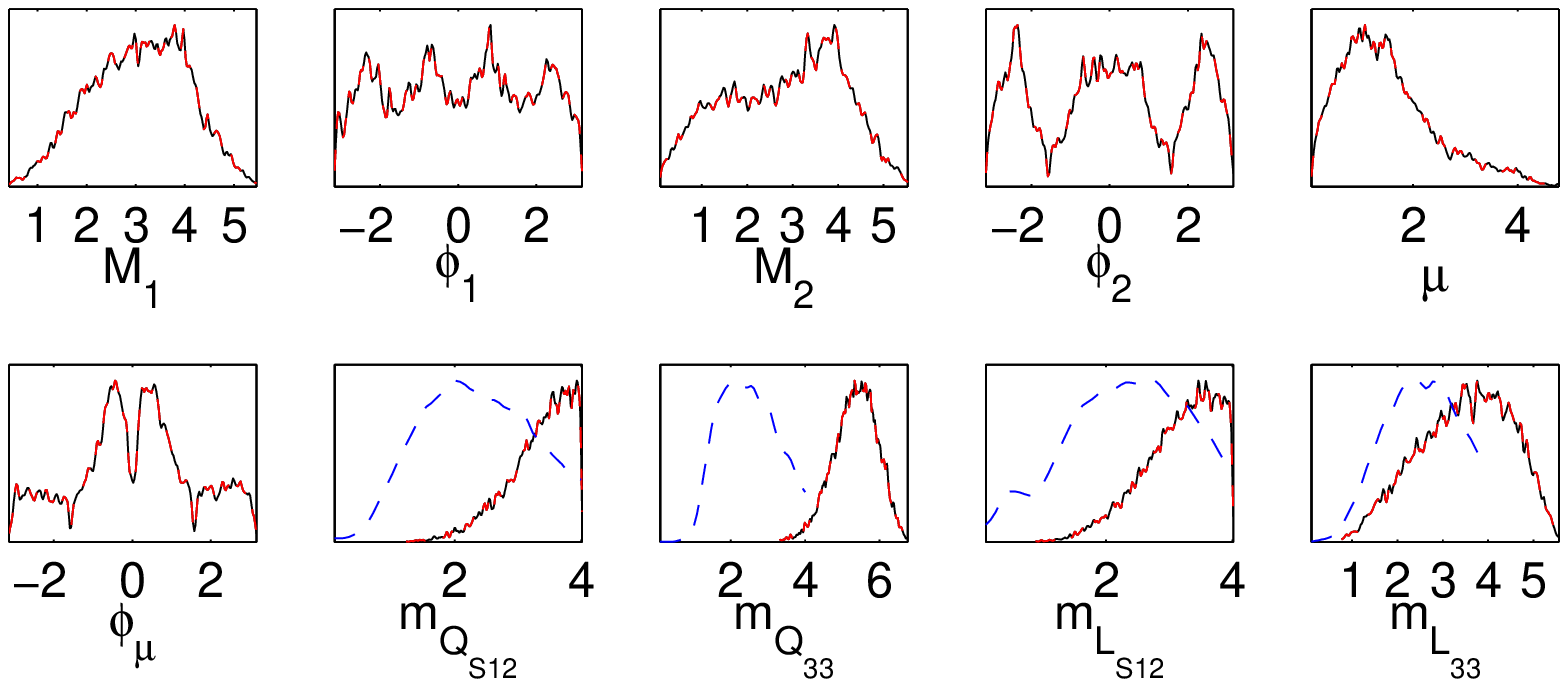,width=.92\textwidth}
\epsfig{file=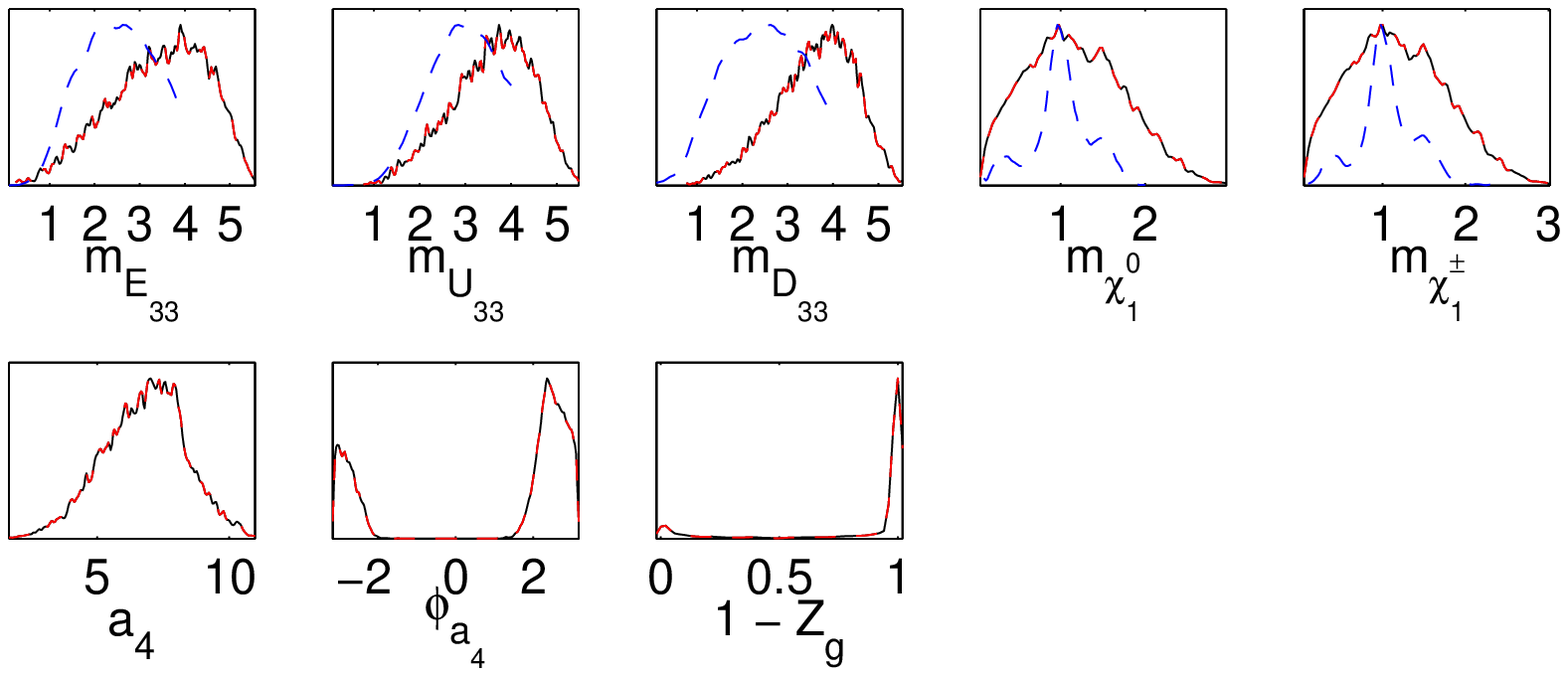,width=.92\textwidth}
\caption{The solid line plots show the posterior distributions for the
  supersymmetry-breaking parameters derived from the base MSSM-30
  parameters shown in Figure~\ref{fig:30pars}. The dashed curves
  represent the corresponding posteriors for the 2008/9 pMSSM
  fits \cite{AbdusSalam:2008uv,AbdusSalam:2009qd} (when these are available) for 
  comparison with the current MSSM-30 fit. The mass parameters are in $\tev$
  and the phases are in radians. The distribution of the neutralino
  composition $1 - Z_g$ as described in the text is also shown.}
\label{fig:par30II}
}

\section{Conclusions and outlook}
\label{sec:concl}
We have implemented the MFV hypothesis'
reparametrisation of the R-parity conserving MSSM as a prescription
for selecting supersymmetry-breaking parameters at various orders, ${\cal
  O}(\lambda^n), n = 1, 2, 3, \ldots$, where $\lambda = \sin
\theta_{CB} = 0.23$ in a Cabibbo mixing angle ($\theta_{CB}$) expansion of
the flavour-violating mass and trilinear coupling terms. This leads to
the construction of the phenomenological MSSM frames, namely MSSM-42
by keeping terms at order ${\mathcal{O}}(\lambda^6)$, MSSM-30 by
keeping terms at order ${\mathcal{O}}(\lambda^4)$, and MSSM-24 by
keeping terms at order ${\mathcal{O}}(\lambda^3)$ with 42, 30 and 24
parameters respectively. The traditional pMSSM cannot be obtained via
this systematic approach because by construction it has 1st-2nd
generation squark mass degeneracies and off-diagonal elements in the
mass terms set to zero {\em by hand}. The MSSM-42, MSSM-30, or MSSM-24
are suitable for fundamental physics studies involving
the usually unavoidable energy- and intensity-frontier effects' interplay.

As a first step within our broader MSSM project, the MSSM-30 is
chosen for going more significantly beyond
the current R-parity conserving MSSM phenomenology constructs. The
MSSM-30 parameters with ${\mathcal O}(\lambda^4) \sim {\cal
  O}(10^{-3})$ coefficients in the MFV basis include the flavour
conserving but CP-violating MSSM phases. We have performed a Bayesian
global fit of the MSSM-30 to experiments data, following the standard
techniques as in Refs.~\cite{AbdusSalam:2008uv,
  AbdusSalam:2009qd}. The data consists of the Higgs boson
mass, the electroweak physics, B-physics, the electric dipole moments
of the leptons and the CDM relic density observables. The posterior
distributions of the 30 parameters are shown in
Figure~\ref{fig:30pars}. The mass term posterior distributions shown
in Figure~\ref{fig:par30II} indicate that the data used favours
multi-TeV 1st/2nd generation and 3rd generation sparticles. 

The preference for smaller/lighter values of $\tan \beta$ and $M_A$
compared to the case for the 2008/9 pMSSM fits
\cite{AbdusSalam:2008uv, AbdusSalam:2009qd} is clear. Their
posterior distributions are approximately prior-independent for
the pMSSM fits~\cite{AbdusSalam:2008uv,
  AbdusSalam:2009qd,2008fit}. This is
also expected to be the case for the MSSM-30 since there is no
feature or observable indicating otherwise, but the study of other
priors is beyond the scope of the present paper\footnote{Notice that
  the comparison with the pMSSM, besides illustrating differences and
  similiarities of the outcomes of the two frameworks may also give us
  further information. Given the detailed analysis available for the
  pMSSM, for instance on the availability of different priors, we may
  also extract some information for the MSSM-30 scenario. Robust
  implications extracted from the pMSSM are not expected to be
  modified from the MSSM-30 analysis and therefore provide further
  information which is a net gain.}. The MSSM-30 
flat-prior fit indicates a $95\%$ Bayesian confidence interval of
$4.5$ to 
$26.9$ for $\tan \beta$ with a mean value of $\langle \tan \beta
\rangle = 13.4 \pm 5.8$. $M_A$ is within the
range of 327.8 to 3803.3 GeV at $95\%$ credible interval with a
mean-value $\langle M_A \rangle = 1751.9 \pm 1024.5
\gev$. These numbers should be taken as indicative to complement the
corresponding posterior distributions shown in
Figure~\ref{fig:30pars}.  It would be interesting to assess the 
effect of the ATLAS and CMS's MSSM Higgs bosons search
results~\cite{Aad:2014vgg,Khachatryan:2014wca} on the MSSM-30
parameter space. This can be done by interpreting the experimental data
within the MSSM-30 such as done in Ref.~\cite{AbdusSalam:2012ir} for
interpreting the supersymmetry results within the pMSSM. It is also 
interesting to find the impact of direct and indirect dark matter
detection data on the MSSM-30. 

Extending our analysis to the more robust MSSM-42
should be achievable in the near future, including a comparison of
different priors in order to extract prior-independent
information. This is a concrete project to follow-up.
This is especially relevant for future studies in search for supersymmetry with
the LHC or some other future collider(s). The power of the Bayesian
approach in determining prior-independent results should be applied
within robust phenomenological frameworks such as the MSSM-30 and
MSSM-42 for this purpose. Its relevance should improve with the
increasing availability of data. Having preference for multi-TeV
supersymmetric particles may also add to the different arguments
supporting higher energy initiatives \cite{FutureColliders} such as a potential 100 TeV
machine\cite{Fowlie:2014awa}.

\acknowledgments
We  acknowledge useful discussions with Liliana
Velasco-Sevilla on the SUSY\_FLAVOUR SLHA-2 inputs and L. Aparicio,
L. Silvestrini and K. Suruliz for discussions. S.S.A. would like to
thank the School of Particles and Accelerators, IPM Tehran, for
hospitality during the finishing phase of this paper. C.B. thanks the
Abdus Salam International Centre for Theoretical Physics for its
hospitality at various points during the completion of this
work. The research leading to these results has received funding from
the European Research Council under the European Union's Seventh
Framework Programme (FP/2007-2013)/ERC Grant Agreement no. 279972
NPFlavour. C.B.~is partially supported by grants from
N.S.E.R.C.~(Canada) and Perimeter Institute for Theoretical
Physics. Research at Perimeter Institute is supported in part by the
Government of Canada through Industry Canada, and by the Province of
Ontario through the Ministry of Research and Information (MRI). This
work was performed using the Darwin Supercomputer of the University of
Cambridge High Performance Computing Service
(http://www.hpc.cam.ac.uk/), provided by Dell Inc. using Strategic 
Research Infrastructure Funding from the Higher Education Funding
Council for England and funding from the Science and Technology
Facilities Council.


\begin{thebibliography}{999}
\bibitem{HHG}
Douglas Adams, {\em The Hitchhiker's Guide to the Galaxy}, Del Rey Publishing 1995.

\bibitem{cMSSM}
See for instance:
S.~P.~Martin,
  %``A Supersymmetry primer,''
  Adv.\ Ser.\ Direct.\ High Energy Phys.\  {\bf 21} (2010) 1
  [hep-ph/9709356];
  %%CITATION = HEP-PH/9709356;%%
  D.~J.~H.~Chung, L.~L.~Everett, G.~L.~Kane, S.~F.~King, J.~D.~Lykken and L.~T.~Wang,
  %``The Soft supersymmetry breaking Lagrangian: Theory and applications,''
  Phys.\ Rept.\  {\bf 407} (2005) 1
  [hep-ph/0312378];
  %%CITATION = HEP-PH/0312378;%%
%\cite{Nath:2010zj}
%\bibitem{Nath:2010zj}
  P.~Nath, B.~D.~Nelson, H.~Davoudiasl, B.~Dutta, D.~Feldman, Z.~Liu, T.~Han and P.~Langacker {\it et al.},
  %``The Hunt for New Physics at the Large Hadron Collider,''
  Nucl.\ Phys.\ Proc.\ Suppl.\  {\bf 200-202} (2010) 185
  [arXiv:1001.2693 [hep-ph]].

\bibitem{MSLED}
  C.~P.~Burgess, J.~Matias and F.~Quevedo,
  %``MSLED: A Minimal supersymmetric large extra dimensions scenario,''
  Nucl.\ Phys.\ B {\bf 706} (2005) 71
  [hep-ph/0404135].


\bibitem{GMSB}
 See for instance:
  G.~F.~Giudice, R.~Rattazzi,
  %``Theories with gauge mediated supersymmetry breaking,''
  Phys.\ Rept.\  {\bf 322 } (1999)  419-499.
  [hep-ph/9801271].

\bibitem{StGMM}
See for instance:
%\bibitem{Ibanez:2012zz}
  L.~E.~Ibanez and A.~M.~Uranga,
  ``String theory and particle physics: An introduction to string phenomenology,''
  Cambridge, UK: Univ. Pr. (2012) 673 p

\bibitem{SimplifiedModels}
%\cite{Alves:2011wf}
%\bibitem{Alves:2011wf}
  D.~Alves {\it et al.}  [LHC New Physics Working Group Collaboration],
  %``Simplified Models for LHC New Physics Searches,''
  J.\ Phys.\ G {\bf 39} (2012) 105005
  [arXiv:1105.2838 [hep-ph]];

\bibitem{NaturalnessModels}
  R.~Kitano and Y.~Nomura,
  %``A Solution to the supersymmetric fine-tuning problem within the MSSM,''
  Phys.\ Lett.\ B {\bf 631} (2005) 58
  [hep-ph/0509039];
  R.~Kitano and Y.~Nomura,
  %``Supersymmetry, naturalness, and signatures at the LHC,''
  Phys.\ Rev.\ D {\bf 73} (2006) 095004
  [hep-ph/0602096];
  M.~Papucci, J.~T.~Ruderman and A.~Weiler,
  %``Natural SUSY Endures,''
  JHEP {\bf 1209} (2012) 035
  [arXiv:1110.6926 [hep-ph]].

\bibitem{pMSSM}
  A.~Djouadi, J.~L.~Kneur and G.~Moultaka,
  %``SuSpect: A Fortran code for the supersymmetric and Higgs particle spectrum in the MSSM,''
  Comput.\ Phys.\ Commun.\  {\bf 176} (2007) 426
  [hep-ph/0211331].

\bibitem{MFV}
  R.~S.~Chivukula and H.~Georgi,
  %``Composite Technicolor Standard Model,''
  Phys.\ Lett.\ B {\bf 188} (1987) 99;
%
  A.~J.~Buras, P.~Gambino, M.~Gorbahn, S.~Jager and L.~Silvestrini,
  %``Universal unitarity triangle and physics beyond the standard model,''
  Phys.\ Lett.\ B {\bf 500} (2001) 161
  [hep-ph/0007085];
%
  G.~D'Ambrosio, G.~F.~Giudice, G.~Isidori and A.~Strumia,
  %``Minimal flavor violation: An Effective field theory approach,''
  Nucl.\ Phys.\ B {\bf 645} (2002) 155
  [hep-ph/0207036];
  %
  A.~J.~Buras,
  %``Minimal flavor violation,''
  Acta Phys.\ Polon.\ B {\bf 34} (2003) 5615
  [hep-ph/0310208].

\bibitem{MFVSUSY1}
  L.~J.~Hall and L.~Randall,
  %``Weak scale effective supersymmetry,''
  Phys.\ Rev.\ Lett.\  {\bf 65} (1990) 2939.

\bibitem{MFVBSM}
  M.~Ciuchini, G.~Degrassi, P.~Gambino and G.~F.~Giudice,
  %``Next-to-leading QCD corrections to B ---> X(s) gamma in supersymmetry,''
  Nucl.\ Phys.\ B {\bf 534} (1998) 3
  [hep-ph/9806308];
  A.~Ali and D.~London,
  %``Profiles of the unitarity triangle and CP violating phases in the standard model and supersymmetric theories,''
  Eur.\ Phys.\ J.\ C {\bf 9} (1999) 687
  [hep-ph/9903535];
  A.~J.~Buras, P.~Gambino, M.~Gorbahn, S.~Jager and L.~Silvestrini,
  %``Universal unitarity triangle and physics beyond the standard model,''
  Phys.\ Lett.\ B {\bf 500} (2001) 161
  [hep-ph/0007085];
    A.~J.~Buras and R.~Fleischer,
  %``Bounds on the unitarity triangle, $\sin$ 2 beta and $K \to$ neutrino anti-neutrino decays in models with minimal flavor violation,''
  Phys.\ Rev.\ D {\bf 64} (2001) 115010
  [hep-ph/0104238];
    C.~Bobeth, T.~Ewerth, F.~Kruger and J.~Urban,
  %``Enhancement of B(anti-B($d$) $\to \mu^{+} \mu^{-)}$ / B(anti-B($s$) $\to \mu^{+} \mu^{-)}$ in the MSSM with minimal flavor violation and large tan beta,''
  Phys.\ Rev.\ D {\bf 66} (2002) 074021
  [hep-ph/0204225];
   A.~J.~Buras,
  %``Relations between $\Delta$ M($s$, $d^{)}$ and B($s$, $d^{)} \to \mu \bar{\mu}$ in models with minimal flavor violation,''
  Phys.\ Lett.\ B {\bf 566} (2003) 115
  [hep-ph/0303060];
     C.~Bobeth, M.~Bona, A.~J.~Buras, T.~Ewerth, M.~Pierini, L.~Silvestrini and A.~Weiler,
  %``Upper bounds on rare K and B decays from minimal flavor violation,''
  Nucl.\ Phys.\ B {\bf 726} (2005) 252
  [hep-ph/0505110];
   V.~Cirigliano, B.~Grinstein, G.~Isidori and M.~B.~Wise,
  %``Minimal flavor violation in the lepton sector,''
  Nucl.\ Phys.\ B {\bf 728} (2005) 121
  [hep-ph/0507001];
  K.~Agashe, M.~Papucci, G.~Perez and D.~Pirjol,
  %``Next to minimal flavor violation,''
  hep-ph/0509117;
  M.~Bona {\it et al.}  [UTfit Collaboration],
  %``The UTfit collaboration report on the status of the unitarity triangle beyond the standard model. I. Model-independent analysis and minimal flavor violation,''
  JHEP {\bf 0603} (2006) 080
  [hep-ph/0509219];
  B.~Grinstein, V.~Cirigliano, G.~Isidori and M.~B.~Wise,
  %``Grand Unification and the Principle of Minimal Flavor Violation,''
  Nucl.\ Phys.\ B {\bf 763} (2007) 35
  [hep-ph/0608123];
  Y.~Grossman, Y.~Nir, J.~Thaler, T.~Volansky and J.~Zupan,
  %``Probing minimal flavor violation at the LHC,''
  Phys.\ Rev.\ D {\bf 76} (2007) 096006
  [arXiv:0706.1845 [hep-ph]];
  A.~L.~Fitzpatrick, G.~Perez and L.~Randall,
  %``Flavor anarchy in a Randall-Sundrum model with 5D minimal flavor violation and a low Kaluza-Klein scale,''
  Phys.\ Rev.\ Lett.\  {\bf 100} (2008) 171604
  [arXiv:0710.1869 [hep-ph]];
  G.~Hiller and Y.~Nir,
  %``Measuring Flavor Mixing with Minimal Flavor Violation at the LHC,''
  JHEP {\bf 0803} (2008) 046
  [arXiv:0802.0916 [hep-ph]];
  P.~Paradisi, M.~Ratz, R.~Schieren and C.~Simonetto,
  %``Running minimal flavor violation,''
  Phys.\ Lett.\ B {\bf 668} (2008) 202
  [arXiv:0805.3989 [hep-ph]];
   T.~Hurth, G.~Isidori, J.~F.~Kamenik and F.~Mescia,
  %``Constraints on New Physics in MFV models: A Model-independent analysis of $\Delta$ F = 1 processes,''
  Nucl.\ Phys.\ B {\bf 808} (2009) 326
  [arXiv:0807.5039 [hep-ph]];
    A.~L.~Kagan, G.~Perez, T.~Volansky and J.~Zupan,
  %``General Minimal Flavor Violation,''
  Phys.\ Rev.\ D {\bf 80} (2009) 076002
  [arXiv:0903.1794 [hep-ph]];
  C.~P.~Burgess, M.~Trott and S.~Zuberi,
  %``Light Octet Scalars, a Heavy Higgs and Minimal Flavour Violation,''
  JHEP {\bf 0909} (2009) 082
  [arXiv:0907.2696 [hep-ph]];
  J.~M.~Arnold, M.~Pospelov, M.~Trott and M.~B.~Wise,
  %``Scalar Representations and Minimal Flavor Violation,''
  JHEP {\bf 1001} (2010) 073
  [arXiv:0911.2225 [hep-ph]];
  J.~M.~Arnold, B.~Fornal and M.~Trott,
  %``Prospects and Constraints for Vector-like MFV Matter at LHC,''
  JHEP {\bf 1008} (2010) 059
  [arXiv:1005.2185 [hep-ph]];
  B.~Batell and M.~Pospelov,
  %``Bs Mixing and Electric Dipole Moments in MFV,''
  Phys.\ Rev.\ D {\bf 82} (2010) 054033
  [arXiv:1006.2127 [hep-ph]];
  B.~Batell, J.~Pradler and M.~Spannowsky,
  %``Dark Matter from Minimal Flavor Violation,''
  JHEP {\bf 1108} (2011) 038
  [arXiv:1105.1781 [hep-ph]];
  M.~Redi,
  %``Composite MFV and Beyond,''
  Eur.\ Phys.\ J.\ C {\bf 72} (2012) 2030
  [arXiv:1203.4220 [hep-ph]];
  W.~Altmannshofer, S.~Gori and G.~D.~Kribs,
  %``A Minimal Flavor Violating 2HDM at the LHC,''
  Phys.\ Rev.\ D {\bf 86} (2012) 115009
  [arXiv:1210.2465 [hep-ph]];
  G.~Krnjaic and D.~Stolarski,
  %``Gauging the Way to MFV,''
  JHEP {\bf 1304} (2013) 064
  [arXiv:1212.4860 [hep-ph]].

\bibitem{MFVSUSY2}
  A.~Ali and E.~Lunghi,
  %``Extended minimal flavor violating MSSM and implications for $B$ physics,''
  Eur.\ Phys.\ J.\ C {\bf 21} (2001) 683
  [hep-ph/0105200];
  S.~Heinemeyer, W.~Hollik, F.~Merz and S.~Penaranda,
  %``Electroweak precision observables in the MSSM with nonminimal flavor violation,''
  Eur.\ Phys.\ J.\ C {\bf 37} (2004) 481
  [hep-ph/0403228];
  G.~Degrassi, P.~Gambino and P.~Slavich,
  %``QCD corrections to radiative B decays in the MSSM with minimal flavor violation,''
  Phys.\ Lett.\ B {\bf 635} (2006) 335
  [hep-ph/0601135];
  W.~Altmannshofer, A.~J.~Buras and D.~Guadagnoli,
  %``The MFV limit of the MSSM for low tan(beta): Meson mixings revisited,''
  JHEP {\bf 0711} (2007) 065
  [hep-ph/0703200];
    G.~Degrassi, P.~Gambino and P.~Slavich,
  %``SusyBSG: A Fortran code for BR[B ---> X(s) gamma] in the MSSM with Minimal Flavor Violation,''
  Comput.\ Phys.\ Commun.\  {\bf 179} (2008) 759
  [arXiv:0712.3265 [hep-ph]];
  M.~Carena, A.~Menon and C.~E.~M.~Wagner,
  %``Minimal Flavor Violation and the Scale of Supersymmetry Breaking,''
  Phys.\ Rev.\ D {\bf 79} (2009) 075025
  [arXiv:0812.3594 [hep-ph]];
  P.~Paradisi and D.~M.~Straub,
  %``The SUSY CP Problem and the MFV Principle,''
  Phys.\ Lett.\ B {\bf 684} (2010) 147
  [arXiv:0906.4551 [hep-ph]];
  J.~Berger, C.~Csaki, Y.~Grossman and B.~Heidenreich,
  %``Mesino Oscillation in MFV SUSY,''
  Eur.\ Phys.\ J.\ C {\bf 73} (2013) 2408
  [arXiv:1209.4645 [hep-ph]].


\bibitem{GIM}
  S.~L.~Glashow, J.~Iliopoulos and L.~Maiani,
  %``Weak Interactions with Lepton-Hadron Symmetry,''
  Phys.\ Rev.\ D {\bf 2} (1970) 1285.


%\cite{Colangelo:2008qp}
\bibitem{Colangelo:2008qp}
  G.~Colangelo, E.~Nikolidakis and C.~Smith,
  %``Supersymmetric models with minimal flavour violation and their running,''
  Eur.\ Phys.\ J.\ C {\bf 59} (2009) 75
  [arXiv:0807.0801 [hep-ph]].

\bibitem{AbdusSalam:2008uv}
  S.~S.~AbdusSalam,
  %``The Full 24-Parameter MSSM Exploration,''
  AIP Conf.\ Proc.\  {\bf 1078 } (2009)  297-299.
  [arXiv:0809.0284]

\bibitem{Berger:2008cq}
  C.~F.~Berger, J.~S.~Gainer, J.~L.~Hewett and T.~G.~Rizzo,
  %``Supersymmetry Without Prejudice,''
  JHEP {\bf 0902} (2009) 023
  [arXiv:0812.0980 [hep-ph]];
  %%CITATION = ARXIV:0812.0980;%%
  M.~W.~Cahill-Rowley, J.~L.~Hewett, A.~Ismail and T.~G.~Rizzo,
  %``More energy, more searches, but the phenomenological MSSM lives on,''
  Phys.\ Rev.\ D {\bf 88} (2013) 3,  035002
  [arXiv:1211.1981 [hep-ph]].
  %%CITATION = ARXIV:1211.1981;%%


%\cite{Ellis:2009di}
\bibitem{Ellis:2009di}
  J.~Ellis, R.~N.~Hodgkinson, J.~S.~Lee and A.~Pilaftsis,
  %``Flavour Geometry and Effective Yukawa Couplings in the MSSM,''
  JHEP {\bf 1002} (2010) 016
  [arXiv:0911.3611 [hep-ph]].


%\cite{AbdusSalam:2009qd}
\bibitem{AbdusSalam:2009qd}
  S.~S.~AbdusSalam, B.~C.~Allanach, F.~Quevedo, F.~Feroz, M.~Hobson,
  %``Fitting the Phenomenological MSSM,''
  Phys.\ Rev.\  {\bf D81 } (2010)  095012.
  [arXiv:0904.2548].

%\cite{MFVRPViol}
\bibitem{MFVRPViol}
  E.~Nikolidakis and C.~Smith,
  %``Minimal Flavor Violation, Seesaw, and R-parity,''
  Phys.\ Rev.\ D {\bf 77} (2008) 015021
  [arXiv:0710.3129 [hep-ph]];
    C.~Csaki, Y.~Grossman and B.~Heidenreich,
  %``MFV SUSY: A Natural Theory for R-Parity Violation,''
  Phys.\ Rev.\ D {\bf 85} (2012) 095009
  [arXiv:1111.1239 [hep-ph]];
   J.~Berger, M.~Perelstein, M.~Saelim and P.~Tanedo,
  %``The Same-Sign Dilepton Signature of RPV/MFV SUSY,''
  JHEP {\bf 1304} (2013) 077
  [arXiv:1302.2146 [hep-ph]];
%\cite{Arcadi:2011ug}
%\bibitem{Arcadi:2011ug}
  G.~Arcadi, L.~Di Luzio and M.~Nardecchia,
  %``Minimal Flavour Violation and Neutrino Masses without R-parity,''
  JHEP {\bf 1205} (2012) 048
  [arXiv:1111.3941 [hep-ph]].


%\cite{Smith:2009hj}
\bibitem{Smith:2009hj}
  C.~Smith,
  %``Minimal flavor violation in supersymmetric theories,''
  Acta Phys.\ Polon.\ Supp.\  {\bf 3} (2010) 53
  [arXiv:0909.4444 [hep-ph]].

%\cite{Feroz:2008wr}
\bibitem{Feroz:2008wr}
  F.~Feroz, B.~C.~Allanach, M.~Hobson, S.~S.~AbdusSalam, R.~Trotta and A.~M.~Weber,
  %``Bayesian Selection of sign(mu) within mSUGRA in Global Fits Including WMAP5 Results,''
  JHEP {\bf 0810} (2008) 064
  [arXiv:0807.4512 [hep-ph]].

%\cite{AbdusSalam:2009tr}
\bibitem{AbdusSalam:2009tr}
  S.~S.~AbdusSalam, B.~C.~Allanach, M.~J.~Dolan, F.~Feroz and M.~P.~Hobson,
  %``Selecting a Model of Supersymmetry Breaking Mediation,''
  Phys.\ Rev.\ D {\bf 80} (2009) 035017
  [arXiv:0906.0957 [hep-ph]].

%\cite{AbdusSalam:2010qp}
\bibitem{AbdusSalam:2010qp}
  S.~S.~AbdusSalam and F.~Quevedo,
  %``Cold Dark Matter Hypotheses in the MSSM,''
  Phys.\ Lett.\ B {\bf 700} (2011) 343
  [arXiv:1009.4308 [hep-ph]].

%\cite{AbdusSalam:2011hd}
\bibitem{AbdusSalam:2011hd}
  S.~S.~AbdusSalam,
  %``Can the LHC rule out the MSSM?,''
  Phys.\ Lett.\ B {\bf 705} (2011) 331
  [arXiv:1106.2317 [hep-ph]].

%\cite{AbdusSalam:2011fc}
\bibitem{AbdusSalam:2011fc}
  S.~S.~AbdusSalam, B.~C.~Allanach, H.~K.~Dreiner, J.~Ellis, U.~Ellwanger, J.~Gunion, S.~Heinemeyer and M.~Kraemer {\it et al.},
  %``Benchmark Models, Planes, Lines and Points for Future SUSY Searches at the LHC,''
  Eur.\ Phys.\ J.\ C {\bf 71} (2011) 1835
  [arXiv:1109.3859 [hep-ph]].

%\cite{AbdusSalam:2012sy}
\bibitem{AbdusSalam:2012sy}
  S.~S.~AbdusSalam and D.~Choudhury,
  %``Higgs boson discovery versus sparticles prediction: Impact on the pMSSM's posterior samples from a Bayesian global fit,''
  UJPA Vol. 2(3) pp.155 - 164, 2014
  [arXiv:1210.3331 [hep-ph]].

%\cite{AbdusSalam:2012ir}
\bibitem{AbdusSalam:2012ir}
  S.~S.~AbdusSalam,
  %``LHC-7 supersymmetry search interpretation within the phenomenological MSSM,''
  Phys.\ Rev.\ D {\bf 87} (2013) 11,  115012
  [arXiv:1211.0999 [hep-ph]].

%\cite{AbdusSalam:2013qba}
\bibitem{AbdusSalam:2013qba}
  S.~S.~AbdusSalam,
  %``Natural stop-mass prediction within MSSM-25,''
  arXiv:1312.7830 [hep-ph].

%\cite{Sekmen:2011cz}
\bibitem{Sekmen:2011cz}
  S.~Sekmen, S.~Kraml, J.~Lykken, F.~Moortgat, S.~Padhi, L.~Pape,
  M.~Pierini and H.~B.~Prosper {\it et al.}, 
  ``Interpreting LHC SUSY searches in the phenomenological MSSM,''
  JHEP {\bf 1202} (2012) 075
  [arXiv:1109.5119 [hep-ph]].

%\cite{Bartl:2001wc}
\bibitem{Bartl:2001wc}
  A.~Bartl, T.~Gajdosik, E.~Lunghi, A.~Masiero, W.~Porod, H.~Stremnitzer and O.~Vives,
  %``General flavor blind MSSM and CP violation,''
  Phys.\ Rev.\ D {\bf 64} (2001) 076009
  [hep-ph/0103324].

%\cite{Ellis:2007kb}
\bibitem{Ellis:2007kb}
  J.~R.~Ellis, J.~S.~Lee and A.~Pilaftsis,
  %``B-Meson Observables in the Maximally CP-Violating MSSM with Minimal Flavour Violation,''
  Phys.\ Rev.\ D {\bf 76} (2007) 115011
  [arXiv:0708.2079 [hep-ph]].

%\cite{Mercolli:2009ns}
\bibitem{Mercolli:2009ns}
  L.~Mercolli and C.~Smith,
  %``EDM constraints on flavored CP-violating phases,''
  Nucl.\ Phys.\ B {\bf 817} (2009) 1
  [arXiv:0902.1949 [hep-ph]].

%\cite{Berger:2013zca}
\bibitem{Berger:2013zca}
  J.~Berger, M.~W.~Cahill-Rowley, D.~Ghosh, J.~L.~Hewett, A.~Ismail and T.~G.~Rizzo,
  %``The CP-violating pMSSM at the Intensity Frontier,''
  arXiv:1309.7653 [hep-ph].


\bibitem{verzo}
  M.~Verzocchi in ``talk at ICHEP 2008'', 2008.

\bibitem{:2005ema}
  {\bf ALEPH} Collaboration,
  %``Precision electroweak measurements on the $Z$ resonance'',
  Phys. Rept. {\bf 427} (2006) 257.

%\cite{Aaij:2012nna}
\bibitem{Aaij:2012nna}
  RAaij {\it et al.}  [LHCb Collaboration],
  %``First Evidence for the Decay $B^0_s \to \mu^+\mu^-$,''
  Phys.\ Rev.\ Lett.\  {\bf 110} (2013) 021801
  [arXiv:1211.2674 [hep-ex]].

      %\cite{Abulencia:2006ze}
    \bibitem{Abulencia:2006ze}
      A.~Abulencia {\it et al.}  [CDF Collaboration],
      %``Observation of B0(s) - anti-B0(s) Oscillations,''
      Phys.\ Rev.\ Lett.\  {\bf 97} (2006) 242003.
      %[hep-ex/0609040].
      %%CITATION = HEP-EX/0609040;%%
      %525 citations counted in INSPIRE as of 18 Apr 2013

      %\cite{Aubert:2004kz}
    \bibitem{Aubert:2004kz}
      B.~Aubert {\it et al.}  [BaBar Collaboration],
      %``Search for the rare leptonic decay $B^- \to \tau^- \bar{\nu}_\tau$,''
      Phys.\ Rev.\ Lett.\  {\bf 95} (2005) 041804.
      %[hep-ex/0407038].
      %%CITATION = HEP-EX/0407038;%%
      %25 citations counted in INSPIRE as of 18 Apr 2013

%\cite{Barberio:2008fa}
\bibitem{Barberio:2008fa}
  E.~Barberio {\it et al.}  [Heavy Flavor Averaging Group
    Collaboration],
  %``Averages of $b-$hadron and $c-$hadron Properties at the End of
  %2007,''
  arXiv:0808.1297 [hep-ex].

      %\cite{Komatsu:2008hk}
    \bibitem{0803.0547}
      E.~Komatsu {\it et al.}  [WMAP Collaboration],
      %``Five-Year Wilkinson Microwave Anisotropy Probe (WMAP) Observations: Cosmological Interpretation,''
      Astrophys.\ J.\ Suppl.\  {\bf 180} (2009) 330.

%\cite{ATLAS:2013mma}
\bibitem{ATLAS:2013mma}
  [ATLAS Collaboration],
  %``Combined measurements of the mass and signal strength of the Higgs-like boson with the ATLAS detector using up to 25 fb$^{-1}$ of proton-proton collision data,''
  ATLAS-CONF-2013-014.


%\cite{CMS:yva}
\bibitem{CMS:yva}
  [CMS Collaboration],
  %``Combination of standard model Higgs boson searches and measurements of the properties of the new boson with a mass near 125 GeV,''
  CMS-PAS-HIG-13-005.

%\cite{Aaij:2011rja}
\bibitem{Aaij:2011rja}
  R.~Aaij {\it et al.}  [LHCb Collaboration],
  %``Search for the rare decays $B^0_s \to \mu^+\mu^-$ and $B^0 \to
  %\mu^+\mu^-$,''
  Phys.\ Lett.\ B {\bf 699} (2011) 330
  [arXiv:1103.2465 [hep-ex]].

%\cite{McNabb:2004tj}
\bibitem{McNabb:2004tj}
  R.~McNabb [Muon g-2 Collaboration],
  %``An Improved limit on the electric dipole moment of the muon,''
  hep-ex/0407008.

      %\cite{Barberio:2007cr}
    \bibitem{Barberio:2007cr}
      E.~Barberio {\it et al.}  [Heavy Flavor Averaging Group (HFAG) Collaboration],
      %``Averages of $b-$hadron properties at the end of 2006,''
      arXiv:0704.3575 [hep-ex].
      %%CITATION = ARXIV:0704.3575;%%
      %374 citations counted in INSPIRE as of 18 Apr 2013

%\cite{Nakamura:2010zzi}
\bibitem{Nakamura:2010zzi}
  K.~Nakamura {\it et al.}  [Particle Data Group Collaboration],
  %``Review of particle physics,''
  J.\ Phys.\ G {\bf 37} (2010) 075021.

%\cite{Regan:2002ta}
\bibitem{Regan:2002ta}
  B.~C.~Regan, E.~D.~Commins, C.~J.~Schmidt and D.~DeMille,
  %``New limit on the electron electric dipole moment,''
  Phys.\ Rev.\ Lett.\  {\bf 88} (2002) 071805.

%\cite{Porod:2003um}
\bibitem{Porod:2003um}
  W.~Porod,
  %``SPheno, a program for calculating supersymmetric spectra, SUSY particle decays and SUSY particle production at e+ e- colliders,''
  Comput.\ Phys.\ Commun.\  {\bf 153} (2003) 275
  [hep-ph/0301101].

%\cite{Porod:2011nf}
\bibitem{Porod:2011nf}
  W.~Porod and F.~Staub,
  %``SPheno 3.1: Extensions including flavour, CP-phases and models beyond the MSSM,''
  Comput.\ Phys.\ Commun.\  {\bf 183} (2012) 2458
  [arXiv:1104.1573 [hep-ph]].

%\cite{Crivellin:2012jv}
\bibitem{Crivellin:2012jv}
  A.~Crivellin, J.~Rosiek, P.~H.~Chankowski, A.~Dedes, S.~Jaeger and P.~Tanedo,
  %``SUSY_FLAVOR v2: A Computational tool for FCNC and CP-violating processes in the MSSM,''
  Comput.\ Phys.\ Commun.\  {\bf 184} (2013) 1004
  [arXiv:1203.5023 [hep-ph]].


%\cite{Camargo-Molina:2013qva}
\bibitem{Camargo-Molina:2013qva}
  J.~E.~Camargo-Molina, B.~O'Leary, W.~Porod and F.~Staub,
  %``$\mathbf{Vevacious}$: A Tool For Finding The Global Minima Of One-Loop Effective Potentials With Many Scalars,''
  Eur.\ Phys.\ J.\ C {\bf 73} (2013) 2588
  [arXiv:1307.1477 [hep-ph]].


%\cite{Camargo-Molina:2014pwa}
\bibitem{Camargo-Molina:2014pwa}
  J.~E.~Camargo-Molina, B.~Garbrecht, B.~O'Leary, W.~Porod and F.~Staub,
  %``Constraining the Natural MSSM through tunneling to color-breaking vacua at zero and non-zero temperature,''
  Phys.\ Lett.\ B {\bf 737} (2014) 156
  [arXiv:1405.7376 [hep-ph]].


%\cite{Allanach:2008qq}
\bibitem{Allanach:2008qq}
  B.~C.~Allanach, C.~Balazs, G.~Belanger, M.~Bernhardt, F.~Boudjema, D.~Choudhury, K.~Desch and U.~Ellwanger {\it et al.},
  %``SUSY Les Houches Accord 2,''
  Comput.\ Phys.\ Commun.\  {\bf 180} (2009) 8
  [arXiv:0801.0045 [hep-ph]].

%\cite{Skands:2003cj}
\bibitem{Skands:2003cj}
  P.~Z.~Skands, B.~C.~Allanach, H.~Baer, C.~Balazs, G.~Belanger, F.~Boudjema, A.~Djouadi and R.~Godbole {\it et al.},
  %``SUSY Les Houches accord: Interfacing SUSY spectrum calculators, decay packages, and event generators,''
  JHEP {\bf 0407} (2004) 036
  [hep-ph/0311123].

%\cite{Belanger:2008sj}
\bibitem{Belanger:2008sj}
  G.~Belanger, F.~Boudjema, A.~Pukhov and A.~Semenov,
  %``Dark matter direct detection rate in a generic model with micrOMEGAs 2.2,''
  Comput.\ Phys.\ Commun.\  {\bf 180} (2009) 747
  [arXiv:0803.2360 [hep-ph]].

%\cite{Heinemeyer:2006px}
\bibitem{Heinemeyer:2006px}
  S.~Heinemeyer, W.~Hollik, D.~Stockinger, A.~M.~Weber and G.~Weiglein,
  %``Precise prediction for M(W) in the MSSM,''
  JHEP {\bf 0608} (2006) 052
  [hep-ph/0604147].

%\cite{Heinemeyer:2007bw}
\bibitem{Heinemeyer:2007bw}
  S.~Heinemeyer, W.~Hollik, A.~M.~Weber and G.~Weiglein,
  %``$Z$ Pole Observables in the MSSM,''
  JHEP {\bf 0804} (2008) 039
  [arXiv:0710.2972 [hep-ph]].

%\cite{Feroz:2007kg}
\bibitem{Feroz:2007kg}
  F.~Feroz and M.~P.~Hobson,
  %``Multimodal nested sampling: an efficient and robust alternative to MCMC methods for astronomical data analysis,''
  Mon.\ Not.\ Roy.\ Astron.\ Soc.\  {\bf 384} (2008) 449
  [arXiv:0704.3704].

%\cite{Feroz:2008xx}
\bibitem{Feroz:2008xx}
  F.~Feroz, M.~P.~Hobson and M.~Bridges,
  %``MultiNest: an efficient and robust Bayesian inference tool for cosmology and particle physics,''
  Mon.\ Not.\ Roy.\ Astron.\ Soc.\  {\bf 398} (2009) 1601
  [arXiv:0809.3437].

\bibitem{Skilling}
J.~{Skilling}, %{\it {Nested Sampling}},
in {\em American Institute of Physics Conference Series}
(R.~{Fischer}, R.~{Preuss}, and U.~V. {Toussaint}, eds.),
pp.~395--405, Nov., 2004.

%\cite{ElKheishen:1992yv}
\bibitem{ElKheishen:1992yv}
  M.~M.~El Kheishen, A.~A.~Aboshousha and A.~A.~Shafik,
  %``Analytic formulas for the neutralino masses and the neutralino mixing matrix,''
  Phys.\ Rev.\ D {\bf 45} (1992) 4345.

%\cite{Feng:1999zg}
\bibitem{Feng:1999zg}
  J.~L.~Feng, K.~T.~Matchev and T.~Moroi,
  %``Focus points and naturalness in supersymmetry,''
  Phys.\ Rev.\ D {\bf 61} (2000) 075005
  [hep-ph/9909334].

%\cite{Aad:2014vgg}
\bibitem{Aad:2014vgg}
  G.~Aad {\it et al.}  [ ATLAS Collaboration],
  %``Search for neutral Higgs bosons of the minimal supersymmetric standard model in pp collisions at $\sqrt{s}$ = 8 TeV with the ATLAS detector,''
  arXiv:1409.6064 [hep-ex].

%\cite{Khachatryan:2014wca}
\bibitem{Khachatryan:2014wca}
  V.~Khachatryan {\it et al.}  [CMS Collaboration],
  %``Search for neutral MSSM Higgs bosons decaying to a pair of tau leptons in pp collisions,''
  arXiv:1408.3316 [hep-ex].

\bibitem{Cohen:1996vb}
  A.~G.~Cohen, D.~B.~Kaplan and A.~E.~Nelson,
  %``The More minimal supersymmetric standard model,''
  Phys.\ Lett.\ B {\bf 388} (1996) 588
  [hep-ph/9607394].

%\cite{Pospelov:2005pr}
\bibitem{Pospelov:2005pr}
  M.~Pospelov and A.~Ritz,
  %``Electric dipole moments as probes of new physics,''
  Annals Phys.\  {\bf 318} (2005) 119
  [hep-ph/0504231].

%\cite{Paradisi:2009ey}
\bibitem{Paradisi:2009ey}
  P.~Paradisi and D.~M.~Straub,
  %``The SUSY CP Problem and the MFV Principle,''
  Phys.\ Lett.\ B {\bf 684} (2010) 147
  [arXiv:0906.4551 [hep-ph]].

%\cite{2008fit}
\bibitem{2008fit}
  S.~AbdusSalam, B.~Allanach, F.~Quevedo, F.~Feroz and M.~Hobson,
  %``The Posterior Sample Points' SLHA Files from the Bayesian Fits of
  %the Phenomenological MSSM,''
  http://dx.doi.org/10.7910/DVN/22742 Harvard Dataverse Network.

%\cite{FutureColliders}
\bibitem{FutureColliders}
The International Linear Collider Technical Design Report - Volumes
1-5, ILC-REPORT-2013-040 (2013); \\
Future Circular Collider (FCC) study web site http://cern.ch/fcc; \\
A Multi-TeV Linear Collider Based on CLIC Technology - CLIC Conceptual
Design Report, CERN-2012-007 (2012).

%\cite{Fowlie:2014awa}
\bibitem{Fowlie:2014awa}
  A.~Fowlie and M.~Raidal,
  %``Prospects for constrained supersymmetry at $\sqrt{s}={33}\,\text {TeV} $ and $\sqrt{s}={100}\,\text {TeV} $ proton-proton super-colliders,''
  Eur.\ Phys.\ J.\ C {\bf 74} (2014) 2948
  [arXiv:1402.5419 [hep-ph]].





\end{thebibliography}
\end{document}